\documentclass[aps,prd,twocolumn,
	superscriptaddress,
	showpacs,preprintnumbers,amsmath,amssymb,nofootinbib,longbibliography
	]{revtex4-1}
\usepackage[
	colorlinks=true,
	pdfstartview=FitV,
	linkcolor=blue,
	citecolor=magenta,
	urlcolor=blue,
	bookmarks=true,
	bookmarksnumbered=true,
	pdftitle={},
	pdfauthor={}  
	]{hyperref}
\usepackage[dvips]{graphicx}
\usepackage{grffile}
\usepackage{braket}
\usepackage{bm,latexsym,amsmath,amssymb,amsfonts,mathrsfs,textcomp}
\usepackage[normalem]{ulem}  
\usepackage{color}
\input{colordvi.tex}

\newcommand\sect[1]{{\it #1.}---}
\newcommand{\qed}{%
	\relax\ifmmode
		\eqno{%
		\setlength{\fboxsep}{2pt}\setlength{\fboxrule}{0.3pt}
		\fbox{\rule[2pt]{0pt}{1ex}}}
	\else
		\begingroup
		\setlength{\fboxsep}{2pt}\setlength{\fboxrule}{0.3pt}
		\hfill\fbox{\rule[2pt]{0pt}{1ex}}
		\endgroup
	\fi}

\newcommand{\Tr}{\mathop{\mathrm{Tr}}}

\newcommand{\average}[1]{\langle#1\rangle}

\newcommand{\averagez}[1]{\langle#1\rangle_z}
\newcommand{\baverage}[1]{\big\langle#1\big\rangle}

\newcommand{\bigaveragez}[1]{\big\langle#1\big\rangle_z}
\newcommand{\Bigaveragez}[1]{\Big\langle#1\Big\rangle_z}
\newcommand{\Biggaveragez}[1]{\Bigg\langle#1\Bigg\rangle_z}
\newcommand{\averageLG}[1]{\langle#1\rangle^{\mathrm{LG}}_{\lambda_t}}
\newcommand{\averageLGzero}[1]{\langle#1\rangle^{\mathrm{LG}}_{\lambda_{t_0}}}

\newcommand{\averageLGG}[1]{\langle#1\rangle^{\mathrm{LG}}_{2\lambda_t}}
\newcommand{\bigaverageLGG}[1]{\big\langle#1\big\rangle^{\mathrm{LG}}_{2\lambda_t}}
\newcommand{\averagetpq}[1]{\langle#1\rangle^{\ell\mathrm{TPQ}}_{\lambda_t}}
\newcommand{\averagetpqzero}[1]{\langle#1\rangle^{\ell\mathrm{TPQ}}_{\lambda_{t_0}}}
\newcommand{\bigaveragetpq}[1]{\big\langle#1\big\rangle^{\ell\mathrm{TPQ}}_{\lambda_t}}

\newcommand{\rme}{\mathrm{e}}
\newcommand{\rmi}{\mathrm{i}}

\newcommand{\hH}{\hat{H}}
\newcommand{\hh}{\hat{h}}

\newcommand{\hK}{\hat{K}}

\newcommand{\hUcal}{\hat{\mathcal{U}}}
\newcommand{\hrho}{\hat{\rho}}

\newcommand{\hSigma}{\hat{\Sigma}}
\newcommand{\hS}{\hat{S}}
\newcommand{\hJ}{\hat{J}}

\newcommand{\hc}{\hat{c}}
\newcommand{\hcurrent}{\hat{\mathcal{J}}}

\newcommand{\hO}{\hat{\mathcal{O}}}

\newcommand{\hA}{\hat{A}}
\newcommand{\hB}{\hat{B}}

\newcommand{\hrhoLG}{\hat{\rho}_{\mathrm{LG}}}

\newcommand{\hOcal}{\hat{\mathcal{O}}}

\newcommand{\hPcal}{\hat{\mathcal{P}}}

\newcommand{\Psf}{\mathsf{P}}
\newcommand{\Tsf}{\mathsf{T}}
\newcommand{\hJsf}{\hat{\mathsf{J}}}

\newcommand{\tildelta}{\widetilde{\delta}}

\newcommand\Ocal{\mathcal{O}}

\newcommand{\diff}{\mathrm{d}}

\renewcommand\Im{\mathop{\mathrm{Im}}}

\newcommand{\bx}{\bm{x}}

\newcommand{\with}{\quad\mathrm{with}\quad}

\newcommand{\cell}{\mathrm{cell}}

\newcommand{\tpq}{\mathrm{TPQ}}

\newcommand{\ltpq}{\ell\mathrm{TPQ}}

\newcommand{\LG}{\mathrm{LG}}
\newcommand{\Prob}{\mathrm{Prob}}

\newcommand{\del}{\partial}
\newcommand{\lp}{\left(}
\newcommand{\rp}{\right)}
\newcommand{\lbb}{\left[}
\newcommand{\rbb}{\right]}
\newcommand{\calH}{\mathcal{H}}
\newcommand{\smrh}{\bar{h}}
\newcommand{\smrJ}{\bar{J}}
\newcommand{\smrO}{\bar{\mathcal{O}}}

\begin{document}
\preprint{RIKEN-QHP-497}

\author{Shoichiro Tsutsui}
\email{shoichiro.tsutsui@riken.jp}
\affiliation{Theoretical Research Division, Nishina Center for Accelerator-Based Science, RIKEN, Wako 351-0198, Japan}

\author{Masaru Hongo}
\email{masaru.hongo@riken.jp}
\affiliation{Department of Physics, University of Illinois, Chicago, Illinois 60607, USA}
\affiliation{RIKEN iTHEMS, RIKEN, Wako 351-0198, Japan}

\author{Shintaro Sato}
\email{shintarosato-phy@ruri.waseda.jp}
\affiliation{Department of Physics, Waseda University, Shinjuku, Tokyo 169-8555, Japan}

\author{Takahiro Sagawa}
\email{sagawa@ap.t.u-tokyo.ac.jp}
\affiliation{
Department of Applied Physics and Quantum-Phase Electronics Center (QPEC), The University of Tokyo, 7-3-1 Hongo, Byunkyo-ku, Tokyo 113-8656, Japan
}

\date{\today}
\title{
Quantum hydrodynamics from local thermal pure states
}

 \begin{abstract}
  We provide a pure state formulation for hydrodynamic dynamics of isolated quantum many-body systems.
  A pure state describing quantum systems in local thermal equilibrium is constructed, which we call a local thermal pure quantum ($\ltpq$) state.
  We show that the thermodynamic functional and the expectation values of local operators (including a real-time correlation function) calculated from the $\ltpq$ state converge to those from a local Gibbs ensemble in the large fluid-cell limit. 
As a numerical demonstration, we investigate a one-dimensional spin chain and observe the hydrodynamic relaxation obeying the Fourier's law.
We further prove the second law of thermodynamics and the quantum fluctuation theorem, which are also validated numerically.
The $\ltpq$ formulation gives a useful theoretical basis to describe the emergent hydrodynamic behavior of quantum many-body systems furnished with a numerical efficiency, being applicable to both the non-relativistic and relativistic regimes. 
 \end{abstract}
\maketitle


\sect{Introduction} 
Nonequilibrium dynamics of isolated quantum many-body systems has been one of the central issues appearing at the crossroad of statistical physics and quantum physics.
In particular, hydrodynamics~\cite{Landau:Fluid,Forster1975,Chaikin2000} is an emergent macroscopic theory to investigate such dynamics under local equilibrium conditions.
Recent experimental technologies have promoted further developments of hydrodynamics theory for isolated quantum systems, ranging from ultracold atoms~\cite{Schafer:2009dj,Cao2011a,Cao2011b,Adams:2012th,Schaefer:2014awa,Elliott2014,Joseph2015,Baird2019,Patel2020} to quark-gluon plasma~\cite{Romatschke:2007mq,Teaney:2009qa,Song:2010mg,Niemi:2011ix,Heinz:2013th,Gale:2013da,Jeon:2015dfa,deSouza:2015ena,Busza:2018rrf,Romatschke2019}.
Furthermore, quantum or classical integrable systems, where conventional hydrodynamics is believed to break down, is now becoming a hot field on the basis of generalized hydrodynamics~\cite{PhysRevX.6.041065,PhysRevLett.117.207201,SciPostPhys.2.2.014,PhysRevLett.121.160603,PhysRevLett.124.140603}.

A recently-developed microscopic derivation of hydrodynamics is based on an extended notion of the statistical ensembles, called the local Gibbs (LG) ensemble both in classical~\cite{Sasa:2013,Mabillard_2020,Mabillard2021} and quantum systems~\cite{Hayata:2015lga,Hongo:2016mqm,Hongo2019,Hongo:2020qpv}.
An advantage of the local ensemble formulation lies in the fact that it does not require the quasi-particle description and thus is applicable to strongly-coupled systems (see also another approach~\cite{Bhattacharyya:2008jc,Rangamani:2009xk,Hartnoll:2009sz,Hubeny:2011hd} based on the holographic principle and the pioneering works~\cite{Nakajima1957,Mori1958,McLennan1960,Luttinger:1964zz,Mori1965,Kawasaki1973,Zubarev} along the line of the linear response theory).
However, under unitary evolutions of isolated quantum systems, a pure state never falls into any mixed state such as the Gibbs ensemble.
Thus, it is still an open problem how macroscopic hydrodynamics emerges under unitary evolutions of pure states as observed in ultracold atoms~\cite{Trotzky2012,Parsons2015,Kaufman2016,Gross2017,Ohl2019,Ebadi:2020ldi}.

In this Letter, we provide a pure state-based derivation of the hydrodynamic equation for isolated quantum many-body systems. 
We introduce a special class of random pure quantum states describing systems in local thermal equilibrium, which we refer to as local thermal pure quantum ($\ltpq$) states.
This is a generalization of thermal pure quantum (TPQ) states~\cite{Bocchieri1959,Hams2000,Popescu2006,Sugiura-Shimizu1,Sugiura-Shimizu2,Monnai2014,Sugiura-Shimizu3,Shimizu2018} randomly sampled from the Hilbert space to reproduce thermal equilibrium behaviors.
We show that $\ltpq$ states have a numerical advantage in computing 
hydrodynamic expectation values~~\cite{Sugiura-Shimizu1,Sugiura-Shimizu2,Sugiura-Shimizu3,Shimizu2018} and give the equivalent results with the conventional method based on the LG ensemble in the large fluid-cell limit.

We establish all the fundamental hydrodynamic behaviors from the $\ltpq$ states,being equipped with a numerical efficiency.
Specifically, we prove the Green-Kubo formula~\cite{Green,Nakano,Kubo}, the second law of thermodynamics and the quantum fluctuation theorem~\cite{Kurchan,Tasaki,RevModPhys.81.1665,RevModPhys.83.771,Sagawa2012,PhysRevLett.119.100601,Funo2018}, as well as the constitutive relation of hydrodynamics itself.
We remark that our formulation applies to both the norelativistic and relativistic regimes, and therefore potentially covers a variety of phenomena from low to high energy scales.
As a concrete demonstration, applying the $\ltpq$-state formulation to a one-dimensional quantum spin chain (see Refs.~\cite{Gemmer2006,Dhar2008,Yonatan2011,Steinigeweg2014a,Lepri2016book,Bernard2016,Bertini2021} and references therein for a review on recent theoretical approaches to thermal transport in low-dimensional systems), we numerically investigate the hydrodynamic relaxation starting from a $\ltpq$ state, and confirm the Fourier's law for thermal conduction.

\sect{Local thermal pure quantum ($\ltpq$) state}
\label{sec:TPQ}
We first introduce the $\ltpq$ state.
Suppose that the system possesses a set of conserved currents
$\hcurrent^\mu_{~a} (t)$ satisfying 
\begin{equation}
 \partial_\mu \hcurrent^\mu_{~a} (t)
  \equiv \partial_0 \hcurrent^0_{~a} (t)
  + \partial_i \hcurrent^i_{~a} (t) = 0,
  \label{eq:Conserv}
\end{equation}
where $\mu,\nu,\cdots$ denote spacetime indices, and $a,b,\cdots$ label
conserved charges such as energy, momentum, and particle numbers%
\footnote{
Summations over the repeated indices for Greek letters ($\mu,\nu,\cdots$)
are assumed throughout the paper.
The repeated indices for Latin letters is assumed to contain the
spatial integral, e.g.
$\lambda^a \hc_a \equiv \int \diff^d x \lambda^a (\bx) \hc_a (\bx)$ where $d$
denotes the spatial dimension.
}.
Throughout the paper, we use the Heisenberg picture, and
Eq.~\eqref{eq:Conserv} gives the equation of motion for conserved charge densities $\hc_a (t) \equiv \hcurrent^0_{~a} (t)$.
Then, we consider the system in local thermal equilibrium, which is described by local thermodynamic parameters $\lambda^a$ conjugate to average charge densities $\average{\hc_a}$.
We assume that each fluid cell contains many degrees of freedom so that thermodynamic parameters are smooth functions of spatial coordinates.
In other words, we identify the current operators $\hcurrent^\mu_{~a}$ in Eq.~\eqref{eq:Conserv} as coarse-graining ones over a fluid-cell volume $V_{\mathrm{cell}}$.

To describe the locally thermalized system, we introduce the $\ltpq$ state at time $t$ by 
\begin{equation}
 \ket{\lambda_t;t}
 \equiv
 \sum_\alpha
 z_\alpha
 \rme^{-\frac{1}{2} \hK[\lambda_t;t]} \ket{\alpha_t} ,
\label{eq:state}
\end{equation}
with $\hK [\lambda_t;t] \equiv \lambda^a(t) \hc_a (t)$,
any orthonormal basis state vector $\ket{\alpha_t}$ at time $t$, and
a complex random variable $z_\alpha \equiv (z_{\alpha}' + \rmi z_{\alpha}'')/\sqrt{2}$
whose real/imaginary parts are taken from the normal distribution.
Note that we distinguish two arguments in $\hK [\lambda_t;t]$; one for thermodynamic parameters $\lambda_t$, whose suffix denotes its configuration $\lambda_t = \lambda(t)$ at time $t$, while the second argument $t$ represents the time argument of charge densities as Heisenberg operators. 
When the system only contains the energy as the conserved charge,
$\hK$ reads $\hK[\lambda_t;t] = \int \diff^d x \beta(t,\bx) \hh (t,\bx)$
with the local inverse temperature $\beta(t,\bx)$ and energy density operator 
$\hh (t,\bx)$, for instance.
The normalization of the $\ltpq$ state is defined by
$Z_{\ltpq} [\lambda_t] \equiv \braket{\lambda_t;t|\lambda_t;t}$,
whose logarithm is identified as a thermodynamic functional.
Using these, we define the entropy-functional operator for the $\ltpq$ state as
\begin{equation}
\hS[\lambda_t;t]
	= \hK [\lambda_t;t ]+ \log Z_{\ltpq} [\lambda_t],
\label{entropy}
\end{equation}
and the average value over the $\ltpq$ state as
\begin{equation}
\averagetpq{\hOcal} \equiv \frac{1}{Z_{\ltpq}[\lambda_t]} 
\braket{\lambda_t;t|\hOcal|\lambda_t;t}.
\end{equation}

\sect{Hydrodynamics and the quantum fluctuation theorem}
\label{sec:TPQ}
Based on the $\ltpq$ state, we derive hydrodynamics and the quantum fluctuation theorem. 
The hydrodynamic equation is regarded as the averaged Eq.~\eqref{eq:Conserv}
over a certain initial density operator $\hrho_0$.
Here, we put a crucial assumption on the initial state: the system is in the $\ltpq$ state at initial time $t_0$ parametrized by $\lambda^a_{t_0}$. 
Starting from this initial $\ltpq$ state, we shall consider the subsequent time-evolution of charge densities described by
\begin{equation}
 \partial_\mu
  \average{\hcurrent^\mu_{~a} (t)}
  = 0
  \with
  \average{\hcurrent^\mu_{~a} (t)}
  \equiv \averagetpqzero{\hcurrent^\mu_{~a} (t)},
  \label{eq:Conserv-Avg}
\end{equation}
for $t \geq t_0$.
To make Eq.~\eqref{eq:Conserv-Avg} a closed set of equations,
we need a constitutive relation, which expresses $\average{\hcurrent^i_{~a} (t)}$ by the dynamical variable $\average{\hc_a(t)} = \average{\hcurrent^0_{~a}(t)}$.
Thus, our problem is to find the constitutive relation.

The vital point here is that we have a solution to this problem if we replace the initial $\ltpq$-state average with the LG-ensemble average over the density operator:
\begin{equation}
 \hrho_{\LG} [\lambda_{t_0};t_0] 
  \equiv \frac{1}{Z_{\LG} [\lambda_{t_0}]}
  \rme^{-\hK[\lambda_{t_0};t_0] },
\end{equation}
where we defined the partition functional for the LG ensemble as $Z_{\LG} [\lambda_{t_0}] \equiv \Tr \rme^{-\hK[\lambda_{t_0};t_0] }$.
We will express the LG average as $\averageLG{\hOcal} \equiv \Tr \big( \hrhoLG [\lambda_t;t] \hOcal \big)$.
Thus, it is enough to show the equivalence between the $\ltpq$ and LG averages for the purpose of deriving hydrodynamics from the $\ltpq$ state.

Following the similar treatment of Refs.~\cite{Sugiura-Shimizu1,Sugiura-Shimizu2,Sugiura-Shimizu3,Shimizu2018}, we can indeed show such equivalence holds as convergence in probability $\xrightarrow{P}$ w.r.t. the random variable $z_a$~\cite{Supp}:
\begin{align}
 Z_{\ltpq} [\lambda_t] &\xrightarrow{P}
 Z_{\LG} [\lambda_t],
 \label{eq:Equiv-Z}
 \\
 \averagetpq{\hOcal} &\xrightarrow{P} \averageLG{\hOcal},
 \label{eq:Equiv-Avg}
\end{align}
where we took the large size limit of all fluid cells $V_{\cell} \to \infty$.
Therefore, the LG-based derivation developed in Refs.~\cite{Hayata:2015lga,Hongo:2016mqm,Hongo2019,Hongo:2020qpv} also provides a derivation of hydrodynamics from the $\ltpq$ state in the large fluid-cell limit.

We here summarize some consequences of Eqs.~\eqref{eq:Equiv-Z}-\eqref{eq:Equiv-Avg}.
Within a first-order derivative expansion supplemented by the Markovian approximation for current-current correlators, the equivalence leads to the constitutive relation 
\begin{equation}
 \begin{split}
  \average{\hcurrent^i_{~a}(t)}
  &= \averagetpq{\hcurrent^i_{~a}(t)}
  + L_{ab}^{ij} (t) \partial_j \lambda^b_t + O(\partial^2),
 \end{split}
 \label{eq:constitutive}
\end{equation}
where transport coefficients $L^{ij}_{ab}$ is given by the Green-Kubo formula with the $\tpq$ state:
\begin{align}
  L_{ab}^{ij} (t)
  &= \beta_t \int_{-\infty}^t \diff t'
  \int \diff^d x' \int_0^1 \diff \tau
  \bigaveragetpq{\tildelta \hcurrent^{i}_{~a,\tau} (x)
  \tildelta \hcurrent^{j}_{~b} (x') } 
 \nonumber \\
 &\simeq \frac{\beta_t}{4V} \int_{-\infty}^{\infty} \diff t'
 \baverage{
 \{\tildelta \hJsf^{i}_{~a} (t),  \tildelta \hJsf^{j}_{~b} (t') \}
 }_{\lambda_t}^{\mathrm{TPQ}},
 \label{eq:Green-Kubo}
\end{align}
where we defined $\average{\hOcal }^{\tpq}_{\lambda_t}$ as a grandcanonical TPQ average parametrized by $\lambda_t$~\cite{Sugiura-Shimizu1,Sugiura-Shimizu2,Sugiura-Shimizu3,Shimizu2018}.
Here we defined a projected operator $\tildelta \hOcal\equiv (1- \hPcal) \delta \hOcal $ using the $\ltpq$ version of Mori's projection~\cite{Mori1965,Kawasaki1973}: 
\begin{equation}
 \hPcal \hOcal (t) \equiv \delta \hc_a (t) \frac{\delta}{\delta c_a(t)}
  \averagetpq{\hOcal},
\end{equation}
together with $\delta \hOcal (t) \equiv \hOcal (t) - \averagetpq{\hOcal (t)}$ 
and $\hOcal_\tau (t)
\equiv \rme^{\hK [\lambda_t;t] \tau} \hOcal (t) \rme^{-\hK [\lambda_t;t]\tau}$.
In the second line of Eq.~\eqref{eq:Green-Kubo}, we neglected the higher-order derivative correction and defined total currents $\hJsf^{i}_{~a} (t) \equiv \int \diff^d x \hcurrent^i_{~a} (t,\bx)$ with the anti-commutator $\{\hA,\hB\} \equiv \hA \hB + \hB \hA$ and total volume $V \equiv \int \diff^d x$.
Note that the constitutive relation~\eqref{eq:constitutive} is expressed by 
the parameter $\lambda_t^a = \{\beta_t, \cdots\}$ at time $t$, which is defined by the matching condition $\average{\hc_a (t)} = \averagetpq{\hc_a (t)}$.
Thanks to one-to-one correspondence between $\lambda^a_t$ and $\average{\hc_a (t)}$ (apart from the first-order phase transition),
Eq.~\eqref{eq:constitutive} indeed expresses $\average{\hcurrent^i_{~a}(t)}$ by $\average{\hc_a (t)}$.
Then, once we evaluate $\averagetpq{\hcurrent^i_{~a}(t)}$ following from the thermodynamic functional $\log Z_{\ltpq} [\lambda_t]$ (see e.g.,~\cite{Hayata:2015lga,Hongo:2016mqm,Hongo2019,Hongo:2020qpv}) and the Green-Kubo formula~\eqref{eq:Green-Kubo}, Eq.~\eqref{eq:constitutive} closes the averaged conservation law \eqref{eq:Conserv-Avg}, which completes an $\ltpq$-based derivation of the first-order hydrodynamics.

In addition to hydrodynamic equations, we can also show the quantum fluctuation theorem and the second law of thermodynamics for the $\ltpq$ state within a small error.
For that purpose, let us introduce the generating function of the total entropy production from initial time $t_0$ to arbitrary time $t(>t_0)$ with forward/backward evolution:
\begin{align}
G_F^{\ltpq} (z) &= 
\braket{
	\hUcal^\dagger(t)
	\rme^{\rmi z \hS [\lambda_t;t_0]}
	\hUcal(t)
	\rme^{-\rmi z \hS [\lambda_{t_0};t_0]}
}_{\lambda_{t_0}}^{\ltpq},
\label{eq:GF}
\\
G_B^{\ltpq} (z) &= 
\braket{
	\widetilde{\mathcal{U}}(t)
	\rme^{\rmi z \hS [\tilde{\lambda}_{t_0};t]}
	\widetilde{\mathcal{U}}^\dagger(t)
	\rme^{-\rmi z \hS [\tilde{\lambda}_{t};t]}
}_{\tilde{\lambda}_{t}}^{\ltpq},
\label{eq:GB}
\end{align}
where $\hUcal(t)$ is the time evolution operator.
By using a combined parity and time-reversal $\Theta \equiv \Psf \Tsf$,
we also defined $\Psf \Tsf$-transformed time evolution operator and parameter by $\widetilde{\mathcal{U}}(t) \equiv \Theta \hUcal(t) \Theta^{-1}$ and 
$\tilde{\lambda}_{t} \equiv \epsilon^a \lambda^a(t,-\bx)$ with $\epsilon^a$ being a $\Psf \Tsf$-eigenvalue of charge densities $\hc_a$, respectively.
We can prove the second law of thermodynamics and
quantum fluctuation theorem as~\cite{Supp}
\begin{align}
 \average{\hS[\lambda_t;t] - \hS[\lambda_{t_0};t_0]}
 &\geq O(\rme^{- A_1 V_{\cell}}),
 \label{eq:SecondLaw}
 \\
 \left|
  G_F^{\ltpq} (z) - G^{\ltpq}_B (-z+\rmi)
 \right|
 &= O(\rme^{- A_2 V_{\cell}}),
 \label{eq:FT}
\end{align}
where $A_1$ and $A_2$ denote certain numerical factors.
Note that the violation of both the fluctuation theorem and the second law is exponentially small with respect to the volume of fluid-cell $V_{\cell}$,
which is the same as Eqs.~\eqref{eq:Equiv-Z}-\eqref{eq:Equiv-Avg}.

\sect{Application to spin chain}
\label{sec:Application}
We now apply the developed formalism to a one-dimensional nonintegrable lattice half-integer spin system, whose Hamiltonian reads
\begin{equation}
\begin{split}
\hH =
\sum_{n=1}^N 
\big[ 
&J_z \sigma_n^z \sigma_{n+1}^z
+ D ( \sigma_n^z \sigma_{n+1}^x - \sigma_n^x \sigma_{n+1}^z ) \\
&+ \Gamma \sigma_n^x + B \sigma_n^z
\big],
\end{split}
\label{hamiltonian}
\end{equation}
where we impose the periodic boundary condition.
The energy is the only conserved quantity of this system so that we find the energy-current operator satisfies the averaged conservation law
\begin{equation}
 \partial_t \average{\hh (t,n)} 
  + \nabla_x \average{\hJ_E (t,n)}
  = 0,
\end{equation}
over the initial $\ltpq$ state $\ket{\beta_{t_0};t_0}$ with 
$\nabla_x \average{\hJ_E (t,n)} \equiv \average{\hJ_E (t,n+1/2)} - \average{\hJ_E (t,n-1/2)}$.

In accordance with our formal derivation, one can expect 
the long-time/length behavior of the energy current is governed by the hydrodynamic equation.
In the present setup, the constitutive relation within the leading-order derivative expansion leads to the Fourier's law: 
\begin{equation}
 \average{\hJ_E (t,n-1/2)}
 = - \kappa_{\mathrm{Fourier}} (t,n) \nabla_x T (t,n-1/2)
 \label{eq:CR-heat}
\end{equation}
where we defined $\nabla_x T (t,n-1/2) \equiv T (t,n) - T (t,n-1)$ with 
$T (t,n) \equiv 1/\beta (t,n)$.
The proportional coefficient $\kappa_{\mathrm{Fourier}} (t,n) $ is identified as the thermal conductivity.
On the other hand, we can also compute the thermal conductivity from the Green-Kubo formula: 
\begin{equation}
\hspace{-8pt} 
 \kappa (t,n)
 = \frac{\beta (t,n)^2}{4N} \!\!
 \lim_{\tau \to \infty} \!
 \int_{-\tau}^\tau \!\!
 \diff t
 \average{\{\delta \hJsf_E (t), \delta \hJsf_E (t_0)\} }^{\tpq}_{\beta(t,n)},
 \label{eq:GK-heat}
\end{equation}
with the total energy current $\hJsf_E (t) \equiv \sum_{n=1}^N \hJ_E (t,n)$.
Our derivation tells us that  $\kappa_{\mathrm{Fourier}}$ should agree with $\kappa$ in the large fluid-cell limit.
Thus, we can quantitatively evaluate the validity of the hydrodynamic description 
 by comparing the values of $ \kappa_{\mathrm{Fourier}} (t,n)$ and $ \kappa (t,n)$, which are independently evaluated from Eqs.~\eqref{eq:CR-heat}-\eqref{eq:GK-heat}.

A numerical implementation of the $\ltpq$-based simulation is
realized in a similar way with the canonical TPQ state~\cite{Sugiura-Shimizu1,Sugiura-Shimizu2,Sugiura-Shimizu3,Shimizu2018}.
Expanding the exponential operator in Eq.~\eqref{eq:state},
we first prepare the initial $\ltpq$ state as
\begin{equation}
 \begin{split}
  &\ket{\beta_{t_0};t_0}
  \equiv 
  \rme^{-\frac{1}{2} \sum_{m=1}^N \beta (t_0,m) l_m}
  \sum_{k=0}^\infty \frac{1}{2^k k!} \\ 
  &\qquad
  \times 
  \lp \sum_{n=1}^N \beta (t_0,n) \big[ l_n - \hat{h} (t_0,n) \big] \rp^k
  \sum_\alpha z_\alpha \ket{\alpha_{t_0}},
 \end{split}
\end{equation}
where $l_n$ is a certain number larger than the maximum eigenvalue of
$\hh (t_0,n)$, and $\beta (t_0,n)$ provides an initial distribution of the
local temperature.
Note that the position dependence of the local temperature does not allow us to
express the $\ltpq$ state by the microcanonical counterpart, as is the
case for global thermal equilibrium~\cite{Sugiura-Shimizu1,Sugiura-Shimizu2,Sugiura-Shimizu3,Shimizu2018}.
Starting from this initial state, we numerically investigate the time evolution
by solving the Schr\"{o}dinger equation with the Runge-Kutta method~\cite{Supp}.

\begin{figure}[t]
 \centering
 \includegraphics[width=1.0\linewidth]{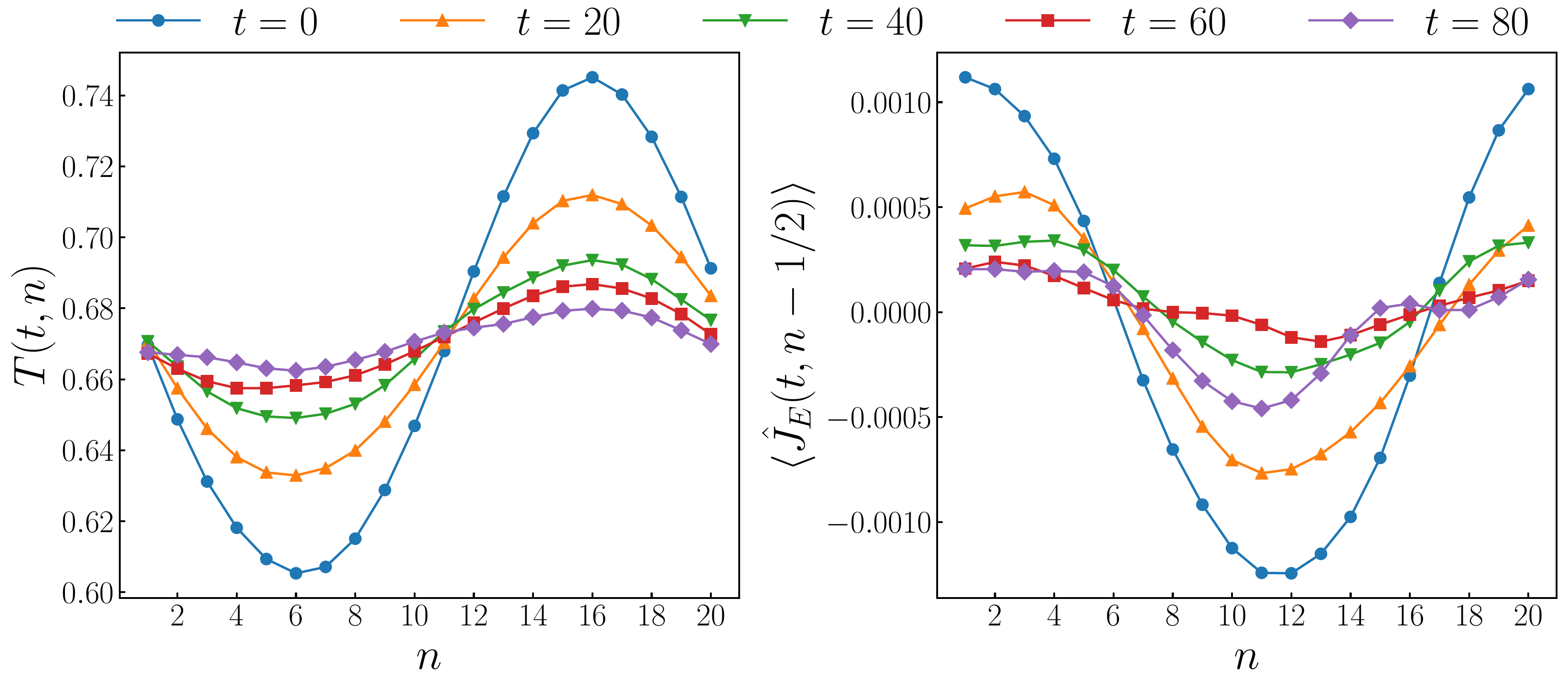}
 \caption{The time-evolutions of (a) local temperature, and (b) energy current 
 of the model \eqref{hamiltonian} with parameters $(J_z,D,\Gamma,B)= (0.2,0.2,0.2,-0.2)$.}
 \label{fig:time-ev}
\end{figure}%

Figure~\ref{fig:time-ev} shows the time-evolution of the local temperature and energy current starting from an inhomogeneous temperature profile: $\beta (t_0,n) = \bar{\beta} + \Delta \beta \sin (2\pi (n-1)/N)$
with $\bar{\beta} = 1.5$ and $\Delta \beta = 0.2$.
Here we imposed the periodic boundary condition with $N=20$.
It is worth emphasizing that we perform the coarse-graining along both spatial and temporal directions: We set the fluid-cell size to contain $5$ spins or so and take the temporal average over $\tau = 10$~\cite{Supp}.
We clearly sees the diffusive relaxation of inhomogeneous temperature (or equivalently energy density).
Furthermore, we can find a tendency of the energy current responding to the temperature gradient, which is compatible with Fourier's law~\eqref{eq:CR-heat}.
Note, however, that we cannot conclude that the system really obeys Eqs.~\eqref{eq:CR-heat}-\eqref{eq:GK-heat} at this point because we have not shown
the thermal conductivity, appearing in Eq.~\eqref{eq:CR-heat} indeed matches with that evaluated from the Green-Kubo formula~\eqref{eq:GK-heat}.

\begin{figure}[b]
 \centering
 \includegraphics[width=1.00\linewidth]{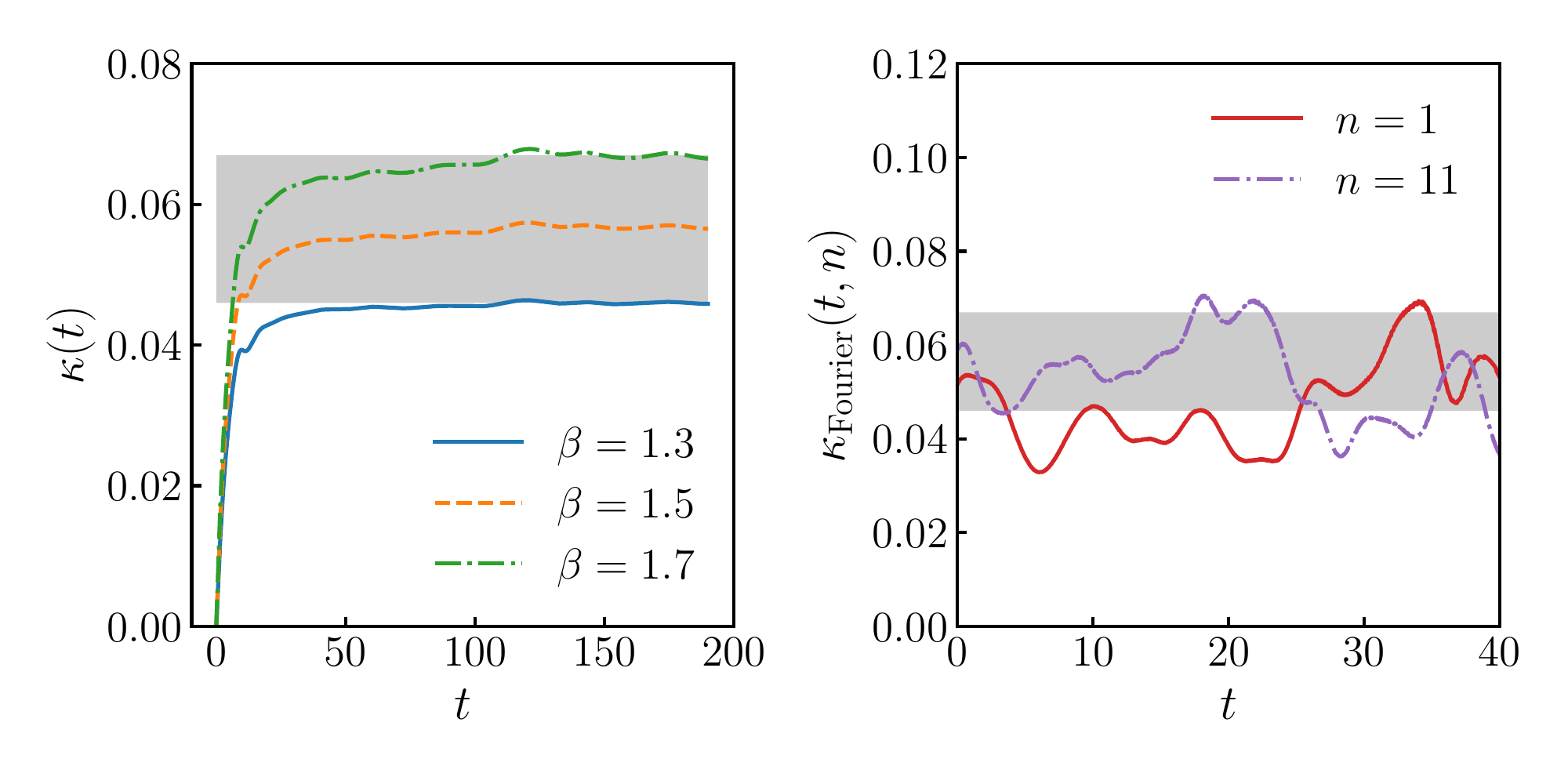}
 \caption{
 Numerical results for 
 (a) Green-Kubo formula \eqref{eq:GK-heat}, and
 (b) comparison of $\kappa_{\mathrm{Fourier}} (t,n)$ with the thermal conductivity evaluated from the Green-Kubo formula. 
 The parameters are the same as in Fig.~\ref{fig:time-ev}.
 }
 \label{fig:heat-conductivity}
\end{figure}

To microscopically confirm the hydrodynamic relaxation, we evaluate the coefficient of Fourier's law \eqref{eq:CR-heat}
and compare it with the thermal conductivity $\kappa (t,n)$ independently  evaluated from the Green-Kubo formula \eqref{eq:GK-heat}.
If the system is really in the hydrodynamic regime, these quantities should coincide with each other.
The result is given in Fig.~\ref{fig:heat-conductivity}.
Taking account of the temperature window, we compute the Green-Kubo formula for three values of temperatures (see shaded regions in Fig.~\ref{fig:heat-conductivity}).
Figure~\ref{fig:heat-conductivity} (b) then indicates that the values of Fourier's law coefficient~\eqref{eq:CR-heat} stay comparable regions with those obtained from the Green-Kubo formula, though they fluctuate depending on temporal and spatial positions.
The agreement may be surprisingly good if we recall our fluid-cell contains just $5$ spins or so, and the deviation is probably the finite-volume effect. 
In short, we conclude that the hydrodynamic description is confirmed to be a good approximation on the basis of the Schr\"odinger equation.

\begin{figure}[t]
 \centering
 \includegraphics[width=0.9\linewidth]{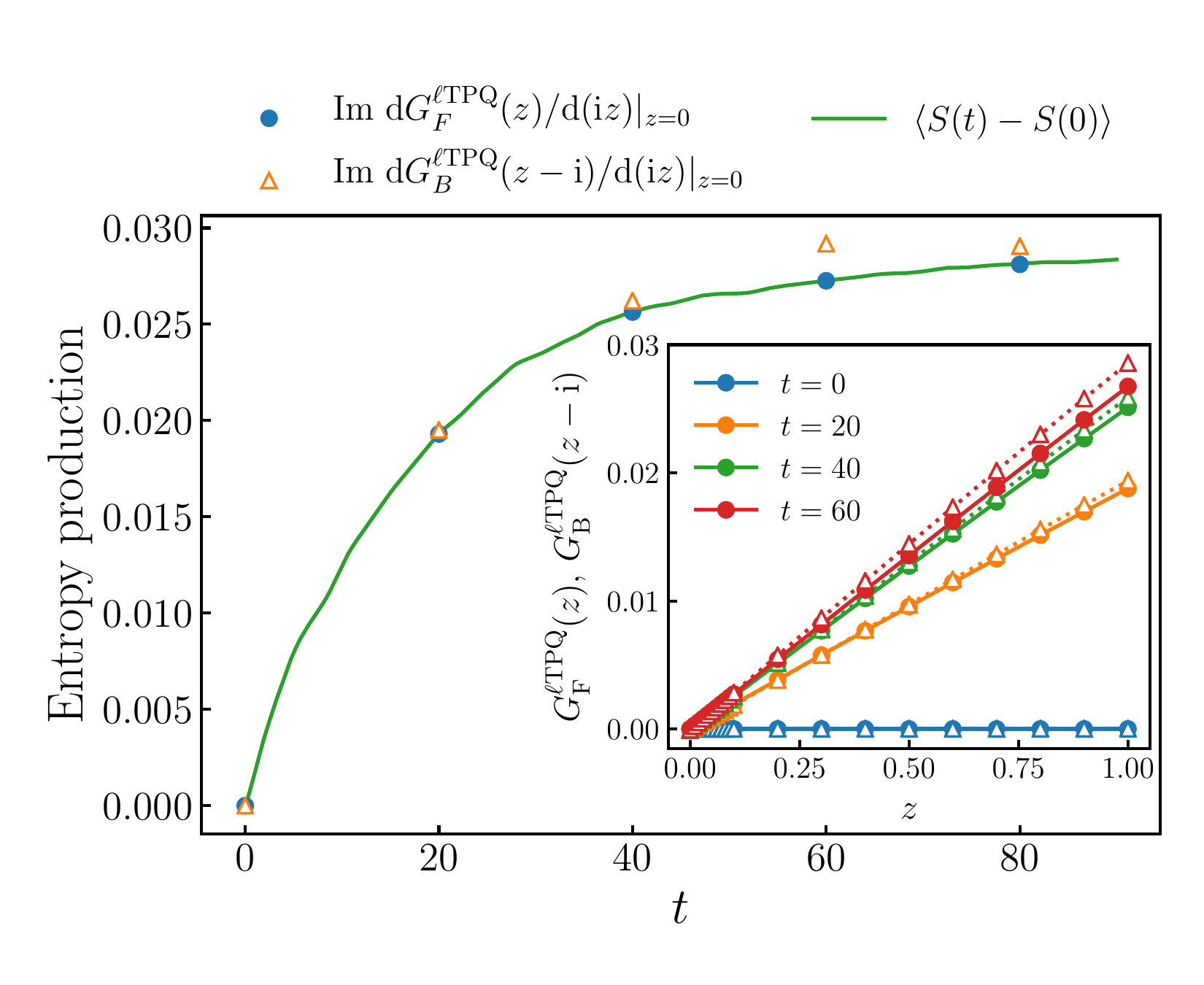}
 \caption{
 A numerical result for the entropy production~\eqref{eq:SecondLaw} [inset: the quantum fluctuation theorem~\eqref{eq:FT}].
 The parameters are the same as in Fig.~\ref{fig:time-ev}.
 }
 \label{fig:second-FT}
\end{figure}

We finally show the result on the second law of thermodynamics and the quantum fluctuation theorem.
Figure~\ref{fig:second-FT} demonstrates the time-dependence of the entropy production (main figure) and the quantum fluctuation theorem (inset).
The solid line in Fig.~\ref{fig:second-FT} for the entropy production is from a direct evaluation, and data are evaluated from the slopes of the generating functions $G_F^{\ltpq} (z)$ and $G_B^{\ltpq} (\rmi - z)$, which are shown in the inset by solid and dotted lines, respectively.
They take positive values in the entire time regions, which confirms the second law of thermodynamics.
Moreover, one sees that $G_F^{\ltpq} (z)$ indeed works as a generating function of the entropy production. 
However, due to uncertainty of the fluctuation theorem coming from the finiteness of the fluid cell as shown in Eq.~\eqref{eq:FT}, $G_B^{\ltpq} (\rmi - z)$ (and its slope) at a later time will deviate from $G_F^{\ltpq} (z)$ (see also the inset of Fig.~\ref{fig:second-FT}).
Although we cannot analytically evaluate dependences of the deviation on 
the time and $z$, this behavior looks similar to those obtained in Ref.~\cite{PhysRevLett.119.100601,Iyoda2021}.

\sect{Concluding remarks}
\label{sec:Summary}
In this Letter, we have formulated hydrodynamics
based on a special class of random pure quantum states, which we call the $\ltpq$ state.
We have shown the equivalence of the $\ltpq$-state formalism
to that of the LG ensemble in the large fluid-cell limit in Eqs.~\eqref{eq:Equiv-Z}-\eqref{eq:Equiv-Avg}, from which 
we have derived the hydrodynamic equations~\eqref{eq:constitutive}-\eqref{eq:Green-Kubo}, the second law of thermodynamics~\eqref{eq:SecondLaw}, and the quantum fluctuation theorem~\eqref{eq:FT}.
Using the $\ltpq$ state, we have performed the numerical simulation of one-dimensional nonintegrable spin chain model.
From a careful analysis, we have numerically confirmed that the observed thermal diffusion allows the hydrodynamic description with the Fourier's law and the Green-Kubo formula (Figs.~\ref{fig:time-ev}-\ref{fig:heat-conductivity}).
The validity of the second law of thermodynamics and quantum fluctuation theorem has also been confirmed (Fig.~\ref{fig:second-FT}).

Let us comment on limitations of the $\ltpq$-state formulation both from the conceptual and practical viewpoints.
First of all, the present formulation employs a $\ltpq$ state mainly motivated by its numerical efficiency. 
However, a $\ltpq$ state is a randomly sampled state to reproduce local thermodynamics for isolated quantum systems. 
Therefore, a $\ltpq$ state is not directly related to the eigenstate thermalization hypothesis (ETH)~\cite{Deutsch1991,Srednicki1994,Rigol2008,Steinigeweg2014b,Kim2014,DAlessio:2016rwt,Dymarsky2018,Yoshizawa2018,Deutsch2018,Noh2020}, as the latter concerns energy eigenstates that are responsible for quantum ergodicity of Hamiltonian dynamics.
We thus believe that it is crucial to construct a pure state formulation based on the ETH in establishing a more solid foundation for hydrodynamics of isolated quantum systems, which is an important future issue.

Another limitation of the $\ltpq$-state formulation is set by the dimension of the Hilbert space of bosonic systems.
Since we do not assume the relativistic or nonrelativistic nature of systems, the $\ltpq$ formulation is, in principle, applicable to any quantum system including the quark-gluon plasma or relativistic/nonrelativistic bose gas.
Although a current computational resource does not allow us to perform the $\ltpq$ simulation for such bosonic systems with a huge degrees of freedom, 
it would be of immediate interest to perform the $\ltpq$ simulation for fermionic systems.

There are also interesting prospects that can be clarified along the line of this Letter.
While we performed a $\ltpq$ simulation for the system with the single conserved quantity, it is interesting to realize the simulation with multiple conserved charges.
In particular, the existence of the momentum density in low-dimensional systems is interersting, because it is expected to drastically change conventional hydrodynamics due to a large hydrodynamic fluctuation~\cite{Lepri2016book}.
Furthemore, it would be an interesting direction to extend our $\ltpq$ framework so as to apply thermal transports for quantum integrable systems described by generalized hydrodynamics~\cite{PhysRevX.6.041065,PhysRevLett.117.207201,SciPostPhys.2.2.014,PhysRevLett.121.160603,PhysRevLett.124.140603}.
We left them as future works (some will be reported in the full paper~\cite{Tsutsui}).

\acknowledgements
The authors thank 
Shunsuke Furukawa, Taiki Morinaga, Keiji Saito, and Shin-ichi Sasa for useful discussions. 
S.\ T.\ was supported by the RIKEN Special Postdoctoral Researchers Program.
M.H. was supported by the US Department of Energy, Office of Science, Office of Nuclear Physics under Award No. DE-FG0201ER41195. 
TS is supported by JSPS KAKENHI Grant Number JP19H05796.
TS is also supported by institute of AI and Beyond of the University of Tokyo.
This work was partially supported by the RIKEN iTHEMS Program, in particular iTHEMS STAMP working group.

\appendix
\bibliography{hydro-tpq}

\pagebreak
\widetext
\begin{center}
\textbf{\large Supplemental Materials:
Quantum hydrodynamics from local thermal pure states}
\end{center}
\setcounter{equation}{0}
\setcounter{figure}{0}
\setcounter{table}{0}
\setcounter{page}{1}
\newcounter{supplementeqcountar}
\newcommand\suppsect[1]{{\it #1.}---}
\renewcommand{\theequation}{S\arabic{equation}}
\renewcommand{\thefigure}{S\arabic{figure}}

\section{Proofs of equivalence in the large fluid-cell limit}

We here give proofs of equivalence between $\ltpq$ formulation and 
the local Gibbs (LG) ensemble method, which includes a proof of 
the second law of thermodynamics and quantum fluctuation theorem in 
the $\ltpq$ formulation.

\subsection{Proof of Eqs.~\eqref{eq:Equiv-Z} and \eqref{eq:Equiv-Avg}}
We here give a proof of Eqs.~\eqref{eq:Equiv-Z}-\eqref{eq:Equiv-Avg},
which is similarly accomplished, as is the case with the global thermal equilibrium~\cite{Sugiura-Shimizu1,Sugiura-Shimizu2,Sugiura-Shimizu3,Shimizu2018}.
For that purpose, we rely on Chebyshev's inequality:
\begin{equation}
 \Prob_z
  \Big[ \left| \Ocal - \averagez{\Ocal} \right| \geq \epsilon \Big]
  \leq \frac{\averagez{\left( \Ocal - \averagez{\Ocal} \right)^2}}{\epsilon^2}
  = \frac{\averagez{\Ocal^2} - \averagez{\Ocal}^2}{\epsilon^2}.
  \label{eq:Cheb}
\end{equation}
Let us first define the average over the random variable $z_{\alpha} = (z_{\alpha}' + \rmi z_{\alpha}'')/\sqrt{2}$ by $\averagez{\Ocal (z)}$.
Since both real and imaginary parts are independently sampled from the
standard normal distribution, they satisfy, e.g.,
\begin{equation}
 \averagez{z_\alpha} = 0, \quad
  \averagez{z_{\alpha_1}^\ast z_{\alpha_2}} = \delta_{{\alpha_1}{\alpha_2}}, \quad
  \averagez{|z_{\alpha}|^4} = 2, \quad
  \averagez{|z_{\alpha_1}|^2 |z_{\alpha_2}|^2} = 1, \quad
  \averagez{z_{\alpha_1}^\ast z_{\alpha_2}^\ast z_{\alpha_3} z_{\alpha_4}} =0,
  \label{eq:zAvg}
\end{equation}
where $\delta_{{\alpha_1}{\alpha_2}}$ denotes the Kronecker delta, and
the last two equations hold for ${\alpha_1} \neq {\alpha_2}$.
With the help of Eqs.~\eqref{eq:zAvg},
we can show the following key identity for a state $\ket{\psi_0} \equiv \sum_{\alpha} z_{\alpha} \ket{{\alpha}}$:
\begin{equation}
 \Bigaveragez{
  \left( \bra{\psi_0} \hA \ket{\psi_0} -
  \bigaveragez{\bra{\psi_0} \hA \ket{\psi_0}} \right)
  \left( \bra{\psi_0} \hB \ket{\psi_0} -
  \bigaveragez{\bra{\psi_0} \hB \ket{\psi_0}} \right) }
  = \Tr \big( \hA \hB \big),
  \label{eq:identity1}
\end{equation}
where $\hA$ and $\hB$ denote arbitrary operators.

\medskip
\noindent \textbf{\underline{Proof of Eq.~\eqref{eq:Equiv-Z}}}:
In order to show the convergence of the partition functional \eqref{eq:Equiv-Z},
we use Chebyshev's inequality \eqref{eq:Cheb} for
$\Ocal = Z_{\ltpq} [\lambda_t]/Z_{\LG} [\lambda_t] = \braket{\lambda_t;t|\lambda_t;t}/\bigaveragez{\braket{\lambda_t;t|\lambda_t;t}}$.
Recalling the definition of the partition functional for the $\ltpq$ state
and LG ensemble $Z_{\LG} [\lambda_t] \equiv \Tr \rme^{- \hK[\lambda_t;t]}$, we find the identity \eqref{eq:identity1} with
$\hA = \hB = \rme^{-\hK[\lambda_t;t]}$ provides the right-hand-side of Chebyshev's
inequality thanks to
\begin{equation}
 \begin{split}
  \bra{\psi_0} \rme^{-\hK[\lambda_t;t]} \ket{\psi_0}
  &= \braket{\lambda_t;t|\lambda_t;t}
  = Z_{\ltpq} [\lambda_t],
  \\
  \bigaveragez{\bra{\psi_0} \rme^{-\hK[\lambda_t;t]} \ket{\psi_0}}
  &= \bigaveragez{\braket{\lambda_t;t|\lambda_t;t}}
  = Z_{\LG} [\lambda_t].
 \end{split}
\end{equation}
As a consequence, we obtain
\begin{equation}
 \Bigaveragez{
  \left( \frac{\braket{\lambda_t;t|\lambda_t;t}}{\bigaveragez{\braket{\lambda_t;t|\lambda_t;t}}} - 1 \right)^2 }
  = \frac{Z_{\LG} [2\lambda_t]}{(Z_{\LG} [\lambda_t])^2}
  = \exp
  \left[ - 2 \int_\Lambda \diff^d x \beta_t
   \Big( p (\beta_t,\mu_t^a) - p (2\beta_t,\mu_t^a)
   + O (\partial^2) \Big)
  \right],
  \label{eq:del1}
\end{equation}
where, to show the second equality, we used the leading-order expression
of the Massieu-Planck functional $\log Z_{\LG} [\lambda_t]$ for
a parity-preserving fluid obtained in Refs.~\cite{Hayata:2015lga,Hongo:2016mqm,Hongo2019}.
Here $\beta_t = \beta (t,\bx)$ denotes the local inverse temperature conjugate to
the energy density, $\mu_t^a = \mu^a (t,\bx)$ local chemical potentials for charge densities
$\hJ_{~a}^0$ attached to internal symmetries of the system labeled by indices $a$.
We explicitly put the size of the fluid cell $\Lambda$.
Substituting this result into Chebyshev's inequality, we obtain
\begin{equation}
 \Prob_z
  \left[ \left|\frac{Z_{\ltpq} [\lambda_t]}{Z_{\LG} [\lambda_t]} - 1 \right|
  \geq \epsilon \right]
  \leq \frac{1}{\epsilon^2}
  \exp \left[ - 2\int_\Lambda \diff^d x
	\beta_t \Big( p(\beta_t,\mu_t^a) - p (2\beta_t,\mu_t^a)
	+ O (\partial^2) \Big) \right].
\end{equation}
The vital point here is that the pressure of a fluid cell $p(\beta,\mu^a)$
is the decreasing function with respect to $\beta$, and satisfies
$p(\beta,\mu^a) - p(2\beta,\mu^a) > 0$.
Therefore, taking the large size limit of the fluid cell
$V_{\cell} \equiv \int_\Lambda \diff^d x$, we see that the probability
of finding the difference between
$Z_{\ltpq} [\lambda_t]$ and $Z_{\LG} [\lambda_t]$ will be exponentially
small.
This completes the proof of Eq.~\eqref{eq:Equiv-Z}.
\qed

\medskip
\noindent \textbf{\underline{Proof of Eq.~\eqref{eq:Equiv-Avg}}}:
Let us move on to the proof for Eq.~\eqref{eq:Equiv-Avg}.
The average value over $\ltpq$ and LG distribution are
given by
\begin{equation}
 \averagetpq{\hOcal} =
  \frac{\bra{\lambda_t;t} \hOcal \ket{\lambda_t;t}}{\braket{\lambda_t;t|\lambda_t;t}}
  \quad \mathrm{and} \quad
  \averageLG{\hOcal} =
  \frac{\bigaveragez{\bra{\lambda_t;t} \hOcal \ket{\lambda_t;t}}}{\bigaveragez{\braket{\lambda_t;t|\lambda_t;t}}}.
  \label{eq:Avgs}
\end{equation}
Noting $\averageLG{\hOcal} \neq \bigaveragez{\averagetpq{\hOcal}}$,
we divide the proof into two steps:
First, we evaluate the difference between $\averagetpq{\hOcal}$ and
$\bigaveragez{\averagetpq{\hOcal}}$, and second, we evaluate 
the difference between $\bigaveragez{\averagetpq{\hOcal}}$ and
$\averageLG{\hOcal}$.

As a first step, we use Chebyshev's inequality for
$\averagetpq{\hOcal}= \bra{\lambda;t} \hOcal \ket{\lambda_t;t}/\braket{\lambda_t;t|\lambda_t;t}$.
In order to evaluate the right-hand-side of that inequality, we expand it as
\begin{equation}
 \begin{split}
  \Bigaveragez{
  \Big( \averagetpq{\hOcal} - \bigaveragez{\averagetpq{\hOcal}} \Big)^2}
  &=
  \Biggaveragez{
  \left(
  \frac{\bra{\lambda_t;t} \hOcal \ket{\lambda_t;t}}{\braket{\lambda_t;t|\lambda_t;t}}
  -
  \Bigaveragez{\frac{\bra{\lambda_t;t} \hOcal \ket{\lambda_t;t}}{\braket{\lambda_t;t|\lambda_t;t}}} \right)^2 }
  \\
  &= \frac{\bigaveragez{(\delta f)^2}}{(Z_{\LG}[\lambda_t])^2}
  - \frac{2 \averageLG{\hOcal} \bigaveragez{ \delta f \delta g}}{(Z_{\LG}[\lambda_t])^2}
  + \frac{\big( \averageLG{\hOcal} \big)^2 \bigaveragez{(\delta g)^2}}{(Z_{\LG}[\lambda_t])^2}
  + O(\delta^3),
 \end{split}
  \label{eq:del2}
\end{equation}
where we defined
\begin{equation}
 \begin{split}
  \delta f
  &=
  \bra{\lambda_t;t} \hOcal \ket{\lambda_tt;t}
  - \bigaveragez{\bra{\lambda_t;t} \hOcal \ket{\lambda_t;t}},
  \\
  \delta g
  &= \braket{\lambda_t;t|\lambda_t;t}
  - \bigaveragez{\braket{\lambda_t;t|\lambda_t;t}}.
 \end{split}
\end{equation}
Since we have already evaluated $\bigaveragez{(\delta g)^2}$
in Eq.~\eqref{eq:del1},
we only need to evaluate the first two terms in Eq.~\eqref{eq:del2};
$\bigaveragez{(\delta f)^2}$ and $\bigaveragez{\delta f \delta g}$.

Let us first evaluate $\bigaveragez{(\delta f)^2}$.
By the use of the identity \eqref{eq:identity1}
with $\hA = \hB = \rme^{-\frac{1}{2} \hK[\lambda_t;t]} \hOcal \rme^{-\frac{1}{2}\hK[\lambda_t;t]}$, we obtain
\begin{equation}
 \bigaveragez{(\delta f)^2}
  = \Tr 
  \left( 
   \rme^{-\hK[\lambda_t;t]} \hOcal \rme^{-\hK[\lambda_t;t]} \hOcal
  \right).
\end{equation}
Since $\hK[\lambda_t;t]$ is an Hermitian operator, taking trace with
its eigenstate $\ket{n}$ result in
\begin{equation}
 \bigaveragez{(\delta f)^2}
  = \sum_{n,m} \rme^{-K_n[\lambda_t;t]} \rme^{- K_m [\lambda_t;t]}
  \bra{n} \hOcal \ket{m} \bra{m} \hOcal \ket{n}
  \leq
  \sum_{n,m} \rme^{-K_n[\lambda_t;t]} \rme^{- K_m [\lambda_t;t]}
  \bra{n} \hOcal \ket{m} \bra{m} \hOcal \ket{n},
\end{equation}
where $K_n[\lambda_t;t]$ denotes the eigenvalue of $\hK[\lambda_t;t]$,
and we used the completeness for the eigenstate of $\hK[\lambda_t;t]$
to show the first equality.
The inequality results from the inequality of arithmetic and geometric means:
$\sqrt{ab} \leq (a+b)/2$.
The right-hand side of this inequality can be further rewritten as
\begin{equation}
 \sum_{n,m} \rme^{-K_n[\lambda_t;t]} \rme^{- K_m [\lambda_t;t]}
  \bra{n} \hOcal \ket{m} \bra{m} \hOcal \ket{n}
  = Z_{\LG} [2\lambda_t] \averageLGG{\hOcal^2},
\end{equation}
and we now obtain the following inequality for $\bigaveragez{(\delta f)^2}$:
\begin{equation}
 \bigaveragez{(\delta f)^2}
  \leq  Z_{\LG} [2\lambda_t] \averageLGG{\hOcal^2}.
  \label{eq:del3}
\end{equation}
To evaluate $\bigaveragez{\delta f \delta g}$, we
simply apply the identity
\eqref{eq:identity1} with
$\hA = \rme^{-\frac{1}{2} \hK[\lambda_t;t]} \hOcal \rme^{-\frac{1}{2} \hK[\lambda_t;t]}$
and $\hB = \rme^{-\hK [\lambda_t;t]}$:
\begin{equation}
 \bigaveragez{\delta f \delta g}
  = \Tr \left( \rme^{-2 \hK[\lambda_t;t]} \hOcal \right)
  = Z_{\LG} [2\lambda_t] \averageLGG{\hOcal}.
  \label{eq:del4}
\end{equation}
Substituting Eqs.~\eqref{eq:del1}, \eqref{eq:del3} and \eqref{eq:del4} into
Eq.~\eqref{eq:del2}, we eventually obtain an upper bound of the variance as
\begin{equation}
 \begin{split}
  \Bigaveragez{
  \Big( \averagetpq{\hOcal} - \bigaveragez{\averagetpq{\hOcal}} \Big)^2}
  &\leq
  \frac{Z_{\LG} [2\lambda_t]}{(Z_{\LG} [\lambda_t])^2}
  \left[
  \averageLGG{\hOcal^2}
  - 2 \averageLG{\hOcal} \averageLGG{\hOcal}
  + \big( \averageLG{\hOcal} \big)^2
  \right]
  \\
  &= \frac{Z_{\LG} [2\lambda_t]}{(Z_{\LG} [\lambda_t])^2}
  \left[
  \bigaverageLGG{ \big( \hOcal - \averageLGG{\hOcal} \big)^2}
  + \big( \averageLG{\hOcal} - \averageLGG{\hOcal} \big)^2
  \right],
  \label{eq:del5}
 \end{split}
\end{equation}
which results in the following inequality:
\begin{equation}
 \Prob_z
  \left[
   \left| \averagetpq{\hOcal} - \bigaveragez{\averagetpq{\hOcal}} \right|
  \geq \epsilon \right]
  \leq \frac{ \rme^{- 2\int_\Lambda \diff^d x
  \beta_t [ p(\beta_t,\mu_t^a) - p (2\beta_t,\mu_t^a) ]) }}{\epsilon^2}
  \left[
   \bigaverageLGG{ \big( \hOcal - \averageLGG{\hOcal} \big)^2}
   + \big( \averageLG{\hOcal} - \averageLGG{\hOcal} \big)^2
  \right].
\end{equation}
As a consequence, for any operator $\hOcal$ whose two variations
$\bigaverageLGG{ \big( \hOcal - \averageLGG{\hOcal} \big)^2}$
and $\big( \averageLG{\hOcal} - \averageLGG{\hOcal} \big)^2$
do not show an exponentially large behavior with respect to
the volume of fluid cells $V_{\cell}$, the inequality
brings about the following result on convergence:
\begin{equation}
 \averagetpq{\hOcal} \xrightarrow{P} \bigaveragez{\averagetpq{\hOcal}},
\end{equation}
which gives the first step of the proof.

Thanks to Eq.~\eqref{eq:del1} and \eqref{eq:del4},
the second step is mostly finished if we evaluate the average
difference by expanding it as follows:
\begin{equation}
 \begin{split}
  \left|
  \bigaveragez{\averagetpq{\hOcal}} - \averageLG{\hOcal}
  \right|
  &= \frac{1}{(Z_{\LG} [\lambda_t])^2}
  \left|
  - \bigaveragez{\delta f \delta gf}
  + \averageLG{\hOcal} \bigaveragez{(\delta g)^2}
  \right|
  \\
  &= \frac{Z_{\LG} [2\lambda_t]}{(Z_{\LG} [\lambda_t])^2}
  \left|
  \averageLGG{\hOcal} - \averageLG{\hOcal}
   \right|
  \\
  &= \exp \left( - 2\int_\Lambda \diff^d x
  \beta_t [ p(\beta_t,\mu_t^a) - p (2\beta_t,\mu_t^a) ]) \right)
  \left|
  \averageLGG{\hOcal} - \averageLG{\hOcal}
   \right|.
  \label{eq:del5}
 \end{split}
\end{equation}
We thus conclude that any operator $\hOcal$ not showing the exponentially large
fluctuation, the difference between
$\bigaveragez{\averagetpq{\hOcal}}$ and $\averageLG{\hOcal}$ is exponentially
small.
This completes the proof of Eq.~\eqref{eq:Equiv-Avg}.
\qed

\subsection{Proof of the Eqs.~\eqref{eq:SecondLaw} and \eqref{eq:FT}}
Although the second law of thermodynamics follows from the quantum
fluctuation theorem~\eqref{eq:FT}, we provide independent proofs of them.

\medskip
\noindent \textbf{\underline{Proof of Eq.~\eqref{eq:SecondLaw}}}:
The second law of thermodynamics can be easily shown by rewriting
the entropy production as follows:
\begin{equation}
 \begin{split}
  \average{\hSigma [t,t_0;\lambda]}
  &\equiv
  \averagetpqzero{\hS [\lambda_t;t]}
  - \averagetpqzero{\hS [\lambda_{t_0};t_0]} 
  \\
  &=
  \underbrace{\averagetpqzero{\hS [\lambda_t;t]} - \averageLGzero{\hS [\lambda_t;t]}}
  _{(a)}
  - \underbrace{
  \big( \averagetpqzero{\hS [\lambda_{t_0};t_0]} - \averageLGzero{\hS [\lambda_{t_0};t_0]} \big)}_{(b)}
  + \underbrace{\averageLGzero{\hS [\lambda_t;t]} - \averageLGzero{\hS [\lambda_{t_0};t_0]}}
  _{(c)}.
 \end{split}
\end{equation}
Using the equivalence between the $\ltpq$ state and LG ensemble,
we can estimate the first and second terms to be exponentially small:
$(a) = O(\rme^{-B_1 V_{\cell}})$ and $(b) = O(\rme^{-B_2 V_{\cell}})$ with
some constants $B_1$ and $B_2$.
Furthermore, the last term is shown to be positive: $(c) \geq 0$ (See Refs.~\cite{Hayata:2015lga,Hongo2019,Hongo:2020qpv}).
This results in Eq.~\eqref{eq:SecondLaw}.
\qed

\medskip
\noindent \textbf{\underline{Proof of Eq.~\eqref{eq:FT}}}:
To prove the quantum fluctuation theorem~\eqref{eq:FT} for the $\ltpq$ state,
we first recall that for the LG ensemble~\cite{Hongo2019}.
For that purpose, let us introduce the generating function for the entropy production under the forward/backward evolution by
\begin{align}
G_F^{\LG} (z) &\equiv
\Tr\left(
\hat{\rho}_{\LG}[\lambda_{t_0};t_0]
\hUcal^\dagger(t)
\rme^{\rmi z \hS [\lambda_t;t_0]}
\hUcal(t)
\rme^{-\rmi z \hS [\lambda_{t_0};t_0]}
\right),
\label{eq:GF-LG}
\\
G_B^{\LG} (z) &\equiv
\Tr \left(
\hat{\rho}_{\LG}[\tilde{\lambda}_{t}; t]	
\widetilde{\mathcal{U}}(t)
\rme^{\rmi z \hS [\tilde{\lambda}_{t_0};t]}
\widetilde{\mathcal{U}}^\dagger(t)
\rme^{-\rmi z \hS [\tilde{\lambda}_{t};t]}
\right),
\label{eq:GB-LG}
\end{align}
where $\hUcal(t)$ is a time evolution operator.
By using a combined parity and time-reversal $\Theta \equiv \Psf \Tsf$,
we also defined $\Psf \Tsf$-transformed time evolution operator and parameter by $\widetilde{\mathcal{U}}(t) \equiv \Theta \hUcal(t) \Theta^{-1}$ and 
$\tilde{\lambda}_{t} \equiv \epsilon^a \lambda^a(t,-\bx)$ with $\epsilon^a$ being a $\Psf \Tsf$-eigenvalue of charge densities $\hc_a$, respectively.
Here, we introduced the LG distribution for the density operator
$\hrhoLG[\lambda_t;t]$ and the entropy functional operator
$\hS_{\LG} [\lambda_t;t]$ by
\begin{equation}
 \hrhoLG [\lambda_t;t]
  \equiv \rme^{- \hS_{\LG} [\lambda_t;t]}
  \quad \mathrm{and} \quad
  \hS_{\LG} [\lambda_t;t]
  \equiv \hK [\lambda_t;t] + \log Z_{\LG} [\lambda_t].
\end{equation}
The generating function in Eqs.~\eqref{eq:GF-LG} and \eqref{eq:GB-LG} are shown to satisfy the quantum fluctuation theorem~\cite{Hongo2019}:
\begin{equation}
 G^{\LG}_F(z) = G^{\LG}_B(-z+\rmi)
  \label{eq:FTforLG}.
\end{equation}

In order to show the quantum fluctuation theorem for the $\ltpq$ state, we then introduce the reference generating function by
\begin{equation}
 \overline{G}_F (z)
  \equiv
  \Tr \left( \hrhoLG [\lambda_{t_0};t_0]
       \rme^{\rmi z\hS_{\ltpq} [\lambda_t;t]} 
       \rme^{-\rmi z\hS_{\ltpq}[t_0;\lambda_{t_0}]}
  \right) .
\end{equation}
By the use of this reference generating function with the triangle inequality,
we can evaluate the difference between $G^{\ltpq}_F(z)$ and $G^{\LG}_F(z)$ as follows:
\begin{equation}
 \begin{split}
  \left| G^{\ltpq}_F(z) - G^{\LG}_F(z) \right|
  &= \left|
  G^{\ltpq}_F(z) - \overline{G}_F (z) - G^{\LG}_F(z) + \overline{G}_F (z)
  \right|
  \\
  &\leq
  \underbrace{\left| G^{\ltpq}_F(z) - \overline{G}_F (z) \right|}_{(A)}
  + \underbrace{\Big| G^{\LG}_F(z) + \overline{G}_F (z) \Big|}_{(B)}.
  \label{eq:DerivationFT1}
 \end{split}
\end{equation}
Defining
$\Delta \Psi [\lambda_t,\lambda_{t_0}] \equiv \log Z[\lambda_t] - \log Z[\lambda_{t_0}] $
for both $\ltpq$ and LG states, and using the error estimation for
the partition functional, we can estimate the second term in the second line as
\begin{equation}
 \begin{split}
  (B)
  &=
  \left|
  \Tr \left( \hrhoLG [\lambda_{t_0};t_0]
  \rme^{\rmi z \hK [\lambda_t;t]} \rme^{-\rmi z \hK [\lambda_{t_0};t_0]}
  \right)
  \left(
  \rme^{\rmi z \Delta \Psi_{\LG} [\lambda_t,\lambda_{t_0}]}
  - \rme^{\rmi z \Delta \Psi_{\ltpq} [\lambda_t,\lambda_{t_0}]}
  \right)
  \right|
  \\
  &=
  \left|
  \Tr \left( \hrhoLG [\lambda_{t_0};t_0]
  \rme^{\rmi z \hK [\lambda_t;t]} \rme^{-\rmi z \hK [\lambda_{t_0};t_0]}
  \right)
  \rme^{\rmi z \Delta \Psi _{\ltpq} [\lambda_t,\lambda_{t_0}]}
  \left(
  \exp \left[ O(\rme^{- C_1 V_{\cell}}) \right]
  - 1
  \right)
  \right|
  \\
  &= O(\rme^{ - C_1 V_{\cell}}).
 \end{split}
 \label{eq:del6}
\end{equation}
Since the first term in Eq.~\eqref{eq:DerivationFT1} is given by
the difference of the averaged operator over the $\ltpq$ state and LG
distribution, we can evaluate it as
\begin{equation}
 \begin{split}
  (A)
  &= \left[
  \averagetpqzero{
  \rme^{\rmi z \hS_{\ltpq} [\lambda_t;t]} 
  \rme^{-\rmi z \hS_{\ltpq} [\lambda_{t_0};t_0]} }
  - \averageLGzero{
  \rme^{\rmi z \hS_{\ltpq} [\lambda_t;t]} 
  \rme^{-\rmi z \hS_{\ltpq} [\lambda_{t_0};t_0]} }
  \right]
  \\
  &= O (\rme^{-C_2 V_{\cell}}),
  \label{eq:del7}
 \end{split}
\end{equation}
where we assumed that the variation of the averaged operator
does not show the exponentially large behavior.
Substituting Eqs.~\eqref{eq:del6}-\eqref{eq:del7} into
Eq.~\eqref{eq:DerivationFT1}, we obtain
\begin{equation}
 \left| G^{\ltpq}_F(z) - G^{\LG}_F(z) \right|
  \leq O(\rme^{-C_3 V_{\cell}}),
  \label{eq:DerivationFT2}
\end{equation}
which states the equivalence of the generating function for the entropy
production for the forward evolution.
On the other hand, a similar analysis also works for
$G_{B}^{\ltpq} (z) - G_B^{\LG} (z)$, and thus, we have,
\begin{equation}
 \left| G^{\ltpq}_B(-z+\rmi) - G^{\LG}_B(-z+\rmi) \right|
  \leq O(\rme^{-C_4 V_{\cell}}),
  \label{eq:DerivationFT3}
\end{equation}
Therefore, combining Eq.~\eqref{eq:FTforLG} with
Eqs.~\eqref{eq:DerivationFT2}-\eqref{eq:DerivationFT3}, we can evaluate
\begin{equation}
 \begin{split}
  &\left| G^{\ltpq}_F(z) - G^{\ltpq}_B(-z+\rmi) \right|
  \\
  =&
  \left| G^{\ltpq}_F(z) - G_F^{\LG} (z)
  - \Big( G_B^{\ltpq} (-z+\rmi) - G_B^{\LG} (-z+\rmi) \Big)
  + G_F^{\LG} (z) -  G^{\LG}_B(-z+\rmi) \right|
  \\
  \leq&
  \left| G^{\ltpq}_F(z) - G_F^{\LG} (z) \right|
  +\left| G_B^{\ltpq} (-z+\rmi) - G_B^{\LG} (-z+\rmi) \right|
  \\
  =& O(\rme^{-A_2} V_{\cell}),
 \end{split}
\end{equation}
where the inequality in the third line again follows from the triangle
inequality.
This gives the quantum fluctuation theorem for the $\ltpq$ state.
\qed

\setcounter{supplementeqcountar}{\value{equation}}
\section{Detail of numrerical implementation}
\setcounter{equation}{\value{supplementeqcountar}}

We here summarize a detail of our numerical simulation based on the $\ltpq$ formulation such as (i) the concrete definition of the (smeared) energy density/current operator, (ii) how to prepare the $\ltpq$ state and the equation of state, 
and (iii) two ways to evaluate the thermal conductivity, and 
(iv) a numerical confirmation of the fluctuation theorem and the second law of thermodynamics.

\subsection{Energy density/current operator and their smearing}\label{sppl:smearing}

\label{supl: energy current operator}
Let us first write the Hamiltonian in the following form:
\begin{align}
\hH = \sum_n \hh(n), \quad  \hh(n) = \calH_{n, n+1} + \calH_{n},
\end{align}
where the energy density operator $\hh(n)$ is decomposed into 
the sum of the nearest-neighbor term $\calH_{n, n+1}$ and the on-site term $\calH_{n}$.
Then, we shall derive the energy current operator $\hJ_E(n-1/2)$.
In order to determine the form of $\hJ_E(n-1/2)$,
we require that the time evolution of the Hamiltonian density should be written as the discrete divergence of the energy current operator as  
\begin{align}
\frac{\del \hh(t,n)}{\del t} = - \big( \hJ_E(t,n+1/2)- \hJ_E(t,n-1/2) \big).
\end{align}
On the other hand,
the Heisenberg equation for the Hamiltonian density operator reads
\begin{align}
\frac{\del \hh(t,n)}{\del t}
= \rmi \big[ \hH, \hh(t,n) \big]
\end{align}
Comparing these two equations, we identify the energy current as follows:
\begin{align}
 \hJ_E(n-1/2) = \rmi\lbb \calH_{n-1, n}, \calH_{n,n+1}+\calH_n \rbb.
\end{align}
Thus, the concrete form of the energy density operator $\hh (n)$
and energy current operator $\hJ_E(n-1/2)$ for the system defined by the 
Hamiltonian~\eqref{hamiltonian} is given by
\begin{align}
 \hh (n) 
 =& J_z \sigma_n^z \sigma_{n+1}^z
 + D ( \sigma_n^z \sigma_{n+1}^x - \sigma_n^x \sigma_{n+1}^z ) 
 + \Gamma \sigma_n^x + B \sigma_n^z
 \\
 \hJ_E(n-1/2)
 =
 & -2 D^2 \lp \sigma_{n-1}^x \sigma_{n}^y \sigma_{n+1}^z - \sigma_{n-1}^z \sigma_{n}^y \sigma_{n+1}^x \rp
 +4D J_z\sigma_{n-1}^z \sigma_{n}^y \sigma_{n+1}^z  \notag \\
 &-2\Gamma J_z\sigma_{n-1}^z \sigma_{n}^y
 +2D \Gamma \sigma_{n-1}^x \sigma_{n}^y
 +2D B \sigma_{n-1}^z \sigma_{n}^y. 
\end{align}

\begin{figure}[tbh]
	\centering
	\includegraphics[width=0.5\linewidth]{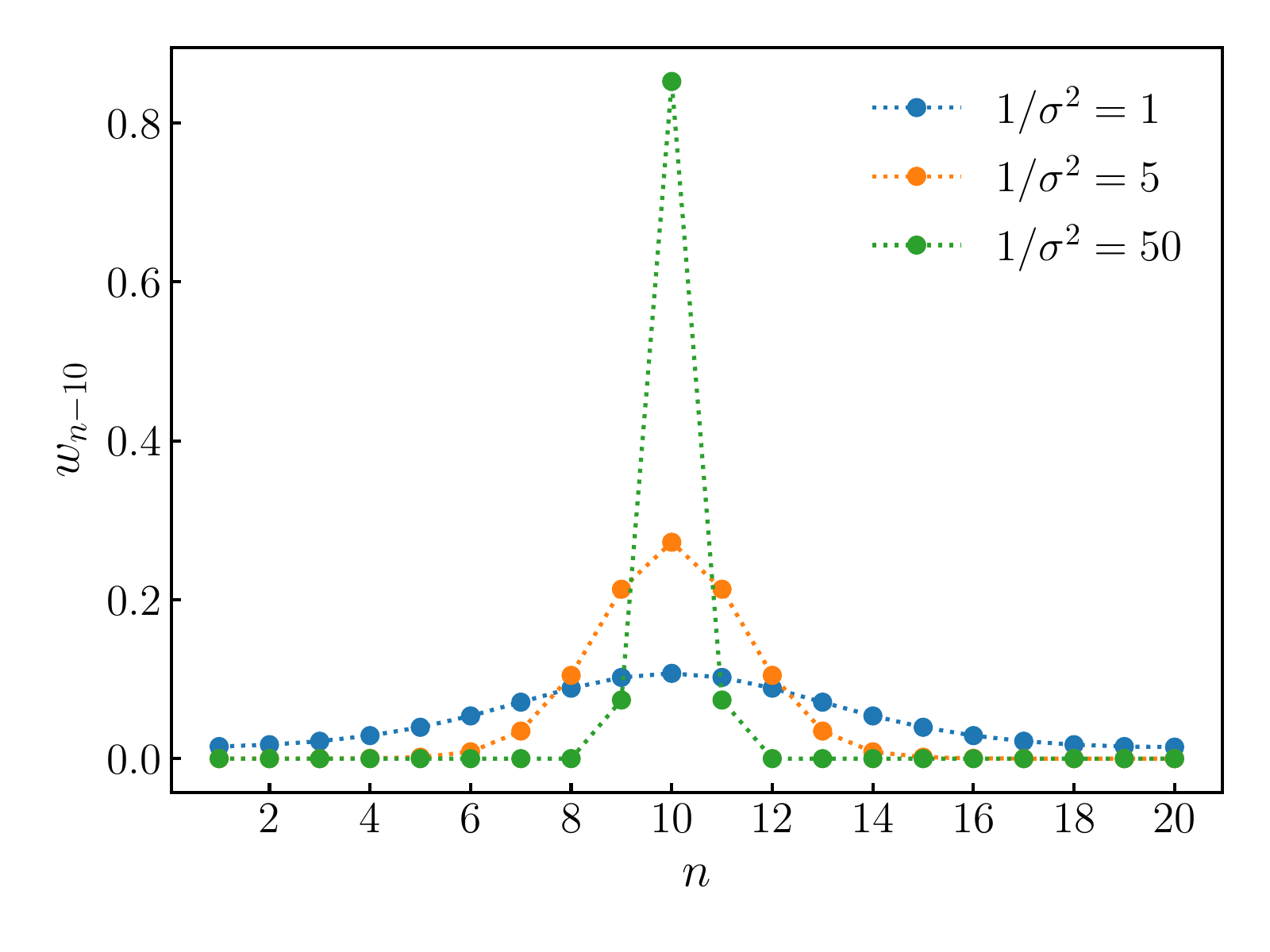}
	\caption{The weight function Eq.~\eqref{weight function}.}
	\label{fig:weight}
\end{figure}

Let us then define a smearing of local operators.
To extract the hydrodynamic behavior, we define a smeared operator of 
an arbitrary local operator $\hO(n)$ by
\begin{align}
\smrO(n) = \sum_{n=1}^N w_{n-m} \hO(m), 
\label{smearing}
\end{align}
with an appropriate weight function $w_{n-m}$ which is a positive and has a peak at $n=m$.
We also require that $w_n$ satisfies the periodic boundary condition $w_{n-N} = w_n$.
This requirement is needed in our setup to define a smeared energy current operator in a consistent manner.
To see this, we apply Eq.~\eqref{smearing} to the local Hamiltonian:
\begin{align}
\frac{\del \smrh(t,n)}{\del t} =
\sum_{m=1}^N  w_{n-m} \frac{\del \hh(t,m)}{\del t}
=
-\lp \sum_{m=1}^N w_{n+1-(m+1/2)}\hJ_E(t,m+1) - \smrJ_E(t,n-1/2) \rp
\end{align}
The first term of the right hand side is rewritten as
\begin{align}
\sum_{m=1}^N w_{n+1-(m+1)}\hJ_E(t,m+1/2) =
\smrJ_E(t,n+1/2) - (w_{n} - w_{n-N})\hJ_E(t,1/2).
\end{align}
Thus, we find that the smeared current operator holds the continuity equation 
\begin{align}
\frac{\del \smrh(t,n)}{\del t} =
-\lp \smrJ_E(t,n+1/2) - \smrJ_E(t,n-1/2) \rp
\end{align}
thanks to the periodicity $w_{n-N} = w_n$.
In our numerical simulations, we choose the weight function $w_n$ as 
\begin{align}
w_n = \frac{1}{\sum_{m=1}^N e^{\frac{1}{\sigma^2} \cos\frac{2\pi m}{N}}} e^{\frac{1}{\sigma^2} \cos\frac{2\pi n}{N}},
\label{weight function}
\end{align}
where $\sigma^2$ is a positive parameter controlling the smearing length.
In the thermodynamic limit with $N\to \infty$, the typical width of this function is given by $1 - I_1(1/\sigma^2)/I_0(1/\sigma^2)$, 
where $I_n(x)$ is the modified Bessel function of the first kind.
In the $1/\sigma^2 \to \infty$ limit for a fixed $N$,
the weight function converges to the Kronecker delta.
For our purpose, the width of the weight function, regarded as the effective size of the fluid cell $\Lambda$, should be taken so that $1 \ll \Lambda \ll N$.
Typical shapes of the weight function is shown in Fig.~\ref{fig:weight}.
For instance, the number of spins in the fluid cell is about 5 when $1/\sigma^2=5$.

\subsection{How to prepare $\ltpq$ state and equation of state}

\paragraph{Construction of the $\ltpq$ state.}

Let us first review a way to implement a canonical TPQ state which is defined by
\begin{align}
\ket{\beta}
\equiv
\exp\lp \frac{-N\beta\hh}{2} \rp \sum_\alpha z_\alpha \ket{\alpha},
\end{align}
where $\hh = \hH/N$ is the Hamiltonian density and $\ket{\alpha}$ is a random vector in the Hilbert space.
We expand  the canonical TPQ state by microcanonical TPQ states as 
\begin{align}
 \ket{\beta}
 = \rme^{-N\beta l/2} \sum_{k=0}^\infty
 R_k \ket{\psi_k}
 \with
 \ket{\psi_k} = \frac{\ket{k}}{\braket{k|k}}, \quad
 \ket{k} \equiv (l-\hh)^k \sum_\alpha z_\alpha \ket{\alpha} 
 \quad \mathrm{and} \quad 
 R_k(\beta) &= \frac{(N\beta/2)^k \braket{k|k}}{k!},
\end{align}
where $l$ is a certain number larger than the maximum eigenvalue of $\hat{h}$.
One can show that $R_k$ has a sharp peak at a certain $k = k^*$, 
whose magnitude is controlled by the inverse temperature~\cite{Sugiura-Shimizu1,Sugiura-Shimizu2,Sugiura-Shimizu3,Shimizu2018}.
Thanks to this property, 
it is justified to truncate the expansion at a certain order $k = k_\text{max}$.
Then, we define $k_\text{max}$ by
\begin{align}
\frac{R_{k_\text{max}}}{\max_k R_{k}} > 10^{-t},
\label{criterion for truncation}
\end{align}
for a given tolerance parameter $t$.
In our numerical simulations, we take $t=20$ and confirm that the expectation values of observables are well converged.

One can generalize the above procedure in a straightforward manner
to the implementation of $\ltpq$ states. 
For that purpose, we first define
\begin{align}
\ket{\{\beta(n)\}}
\equiv
\exp\lp\frac{-\sum_{n=1}^N \beta(n) \hh(n)}{2}\rp \sum_\alpha z_\alpha \ket{\alpha},
\end{align}
where $\hh_n$ is the local Hamiltonian density.
Expanding the exponential operator, we find
\begin{align}
\ket{\{\beta(n)\}}
&=
 \rme^{-\frac{1}{2}\sum_n\beta(n)l_n}
 \exp\lp\frac{\sum_{n=1}^N \beta(n) (l_n -\hh(n))}{2}\rp \sum_\alpha z_\alpha \ket{\alpha} \\
 &=
 \rme^{-\frac{1}{2}\sum_n\beta(n)l_n}
 \sum_k \frac{1}{2^kk!}\lp \sum_{n=1}^N \beta(n) (l_n-\hh(n))\rp^k
 \sum_\alpha z_\alpha \ket{\alpha} \\
 &=
 \rme^{-\frac{1}{2}\sum_n\beta(n)l_n}
 \sum_k \frac{\bar{\beta}^kN^k}{2^kk!}
 \lp \frac{1}{N\bar{\beta}}\sum_{n=1}^N \beta(n) (l_n-\hh(n))\rp^k
 \sum_\alpha z_\alpha \ket{\alpha} \label{expanded lTPQ state},
\end{align}
where $l_n$ is a certain number larger than the maximum eigenvalue of $\hat{h}$,
and $\bar{\beta} = \frac{1}{N}\sum_{n=1}^N \beta(n)$ is the average inverse temperature.
We note that the state appeared in Eq.~\eqref{expanded lTPQ state}
\begin{align}
\ket{k, \{\beta(n)\}} \equiv \lp\frac{1}{N\bar{\beta}}\sum_{n=1}^N \beta(n) (l_n-\hat{h}(n))\rp^k \sum_\alpha z_\alpha \ket{\alpha}
\end{align}
can be regarded as a generalization of the microcanonical TPQ state $\ket{k}$:
Indeed, we can show $\ket{k, \{\beta(n)\}} = \ket{k}$ if $\beta = \beta_1 = \dots = \beta_N$, $l = l_1 = \dots, l_N$.
Then, we obtain the expression the $\ltpq$ state as follows:
\begin{align}
 &\ket{\{\beta(n)\}}
=
e^{-N\beta l/2} \sum_{k=0}^\infty
R_k(\{\beta(n)\}) \ket{\psi_k, \{\beta(n)\}}, \label{lTPQ state practical form} \\
 & \with
 \ket{\psi_k, \{\beta(n)\}} = \frac{\ket{k, \{\beta(n)\}}}{\braket{k, \{\beta(n)\}|k, \{\beta(n)\}}}, \quad
 R_k(\{\beta(n)\}) = \frac{(N\beta/2)^k \braket{k, \{\beta(n)\}|k, \{\beta(n)\}}}{k!}.
\label{lTPQ state practical form2}
\end{align}
We confirmed that $R_k(\{\beta(n)\})$ also has a sharp peak at a certain $k = k^*$ so that we can truncate the summation over $k$ in Eq.~\eqref{lTPQ state practical form} in a same manner as the canonical TPQ states.
Equations \eqref{expanded lTPQ state}-\eqref{lTPQ state practical form2} define  a way to prepare the $\ltpq$ state starting from an arbitrary random vector 
$\ket{a}$ in the Hilbert space.

\paragraph{The energy-temperature relation.}

In the main article, we introduce the local temperature according to
the matching condition $\average{\bar{h}(t,n)} = \averagetpq{\bar{h}(t,n)} $.
Within the derivative expansion, this condition is equivalent to use the thermodynamic relation to determine the local temperature from the energy density.

\begin{figure}[tbh]
	\centering
	\includegraphics[width=0.5\linewidth]{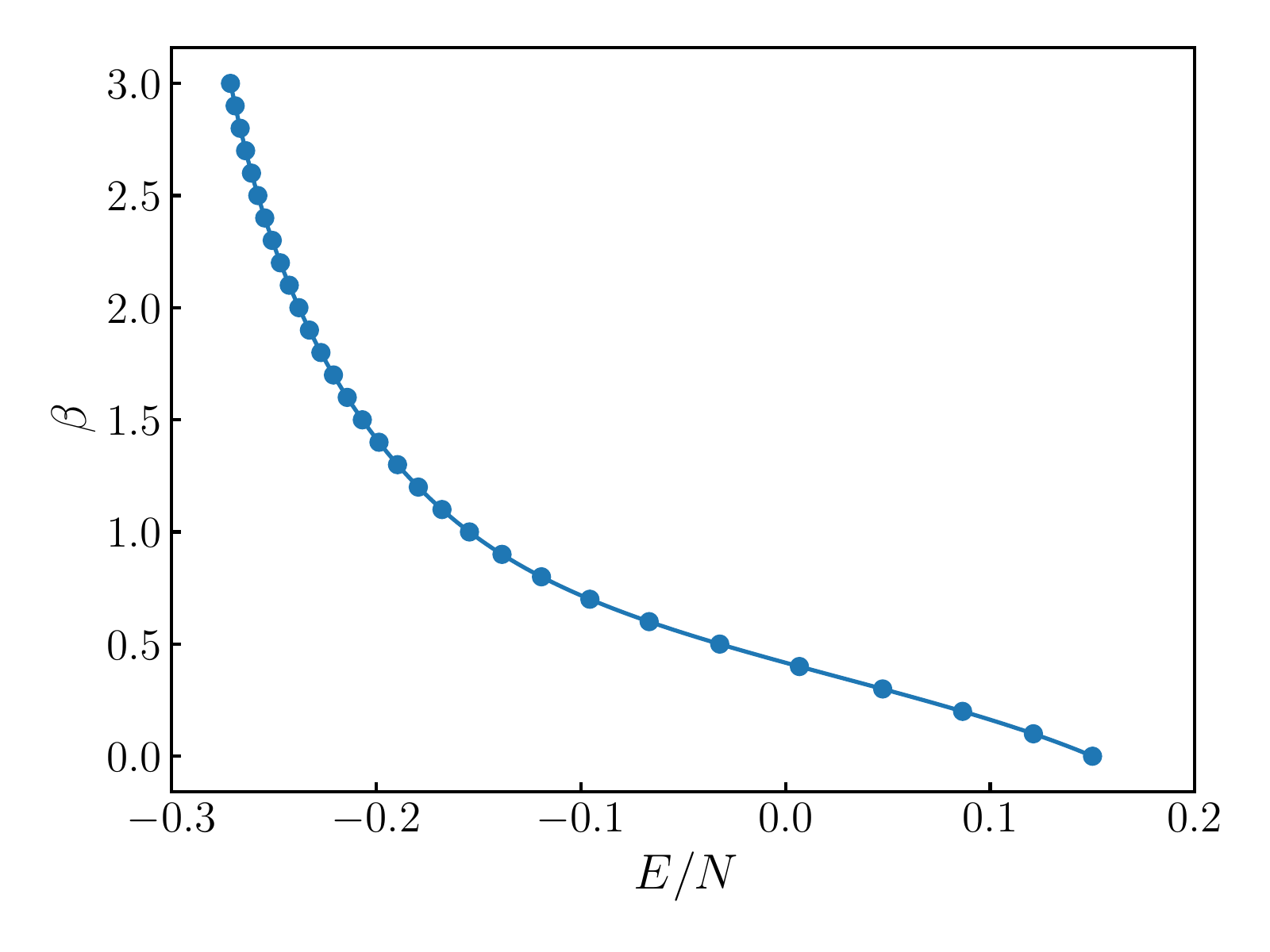}
	\caption{The equation of state for the Hamiltonian Eq.~\eqref{hamiltonian} with the system size $N=20$. The total energy is calculated at $\beta=0.0, 0.1, \cdots, 3.0$ and they are represented by circles. The solid line is the interpolation of them.}
	\label{fig:eos}
\end{figure}

One can use the equation of state as such a thermodynamic relation, which 
enables us to relate the energy density and local temperature.
Using the canonical TPQ states, the total energy of the system is calculated as 
\begin{align}
    E \equiv \frac{\braket{\beta|\hH|\beta}}{\braket{\beta|\beta}}.
\end{align}
This relation gives the one-to-one correspondence between the energy and the inverse temperature as long as first order phase transitions do not occur.
We calculate the total energy for several inverse temperatures $\beta=0.0, 0.1, \cdots, 3.0$ for the system size $N=20$.
The coupling constants are set to $J_z=\Gamma=-B=D=0.2$. 
The results is shown in Fig.~\ref{fig:eos}.

\subsection{Two ways to evaluate thermal conductivity}

\paragraph{Thermal conductivity from the Fourier law.}
While the formal arguments rely on the Heisenberg picture,
we perform numerical simulations based on the Schr\"{o}dinger picture.
The expectation value of an operator $\hat{O}$ at time $t$ is calculated by
\begin{align}
&\braket{\hat{O}(t)} = \frac{\braket{\psi(t)|\hat{O}|\psi(t)}}{\braket{\psi(0)|\psi(0)}},  \\
\rmi\frac{\del}{\del t} \ket{\psi(t)} &= \hH \ket{\psi(t)}, \quad \ket{\psi(0)} = \ket{\{\beta(n)\}}. 
\end{align}
Unless otherwise noted,
the number of spins is taken as $N=20$,
and the coupling constants are $J_z=\Gamma=-B=D=0.2$ so that the system is nonintegrable.
We set the initial temperature profile of the $\ltpq$ state as
\begin{align}
\beta(n) = \bar{\beta} + \Delta \beta \sin (2\pi (n-1)/N), \quad n=1,\dots, N,
\label{eq: initial profile}
\end{align}
with $\bar{\beta} = 1.5$ and $\Delta \beta = 0.2$.
We numerically solve the Schr\"{o}dinger equation by the fourth order Runge-Kutta method with a time slice $\Delta t = 0.01$,
and calculated the local energy and the energy current.
The smearing is also performed by using the weight function given in Eq.~\eqref{weight function} with $1/\sigma^2=5$.
The local temperature is estimated from the local energy through the relation in equilibrium shown in Fig.~\ref{fig:eos} assuming that derivative corrections are negligible.

\begin{figure}[tb]
	\centering
	\includegraphics[width=0.325\linewidth]{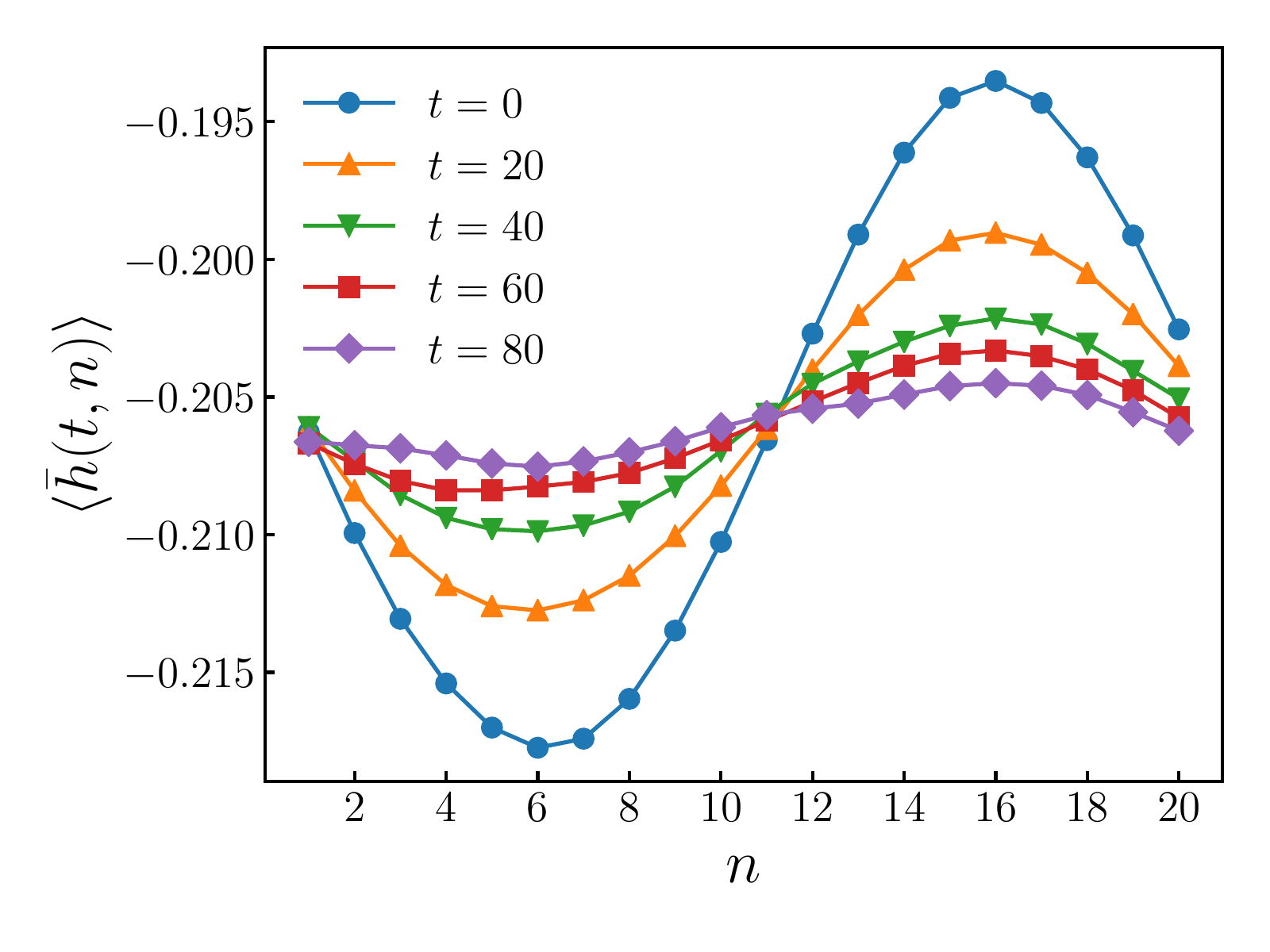}
	\includegraphics[width=0.325\linewidth]{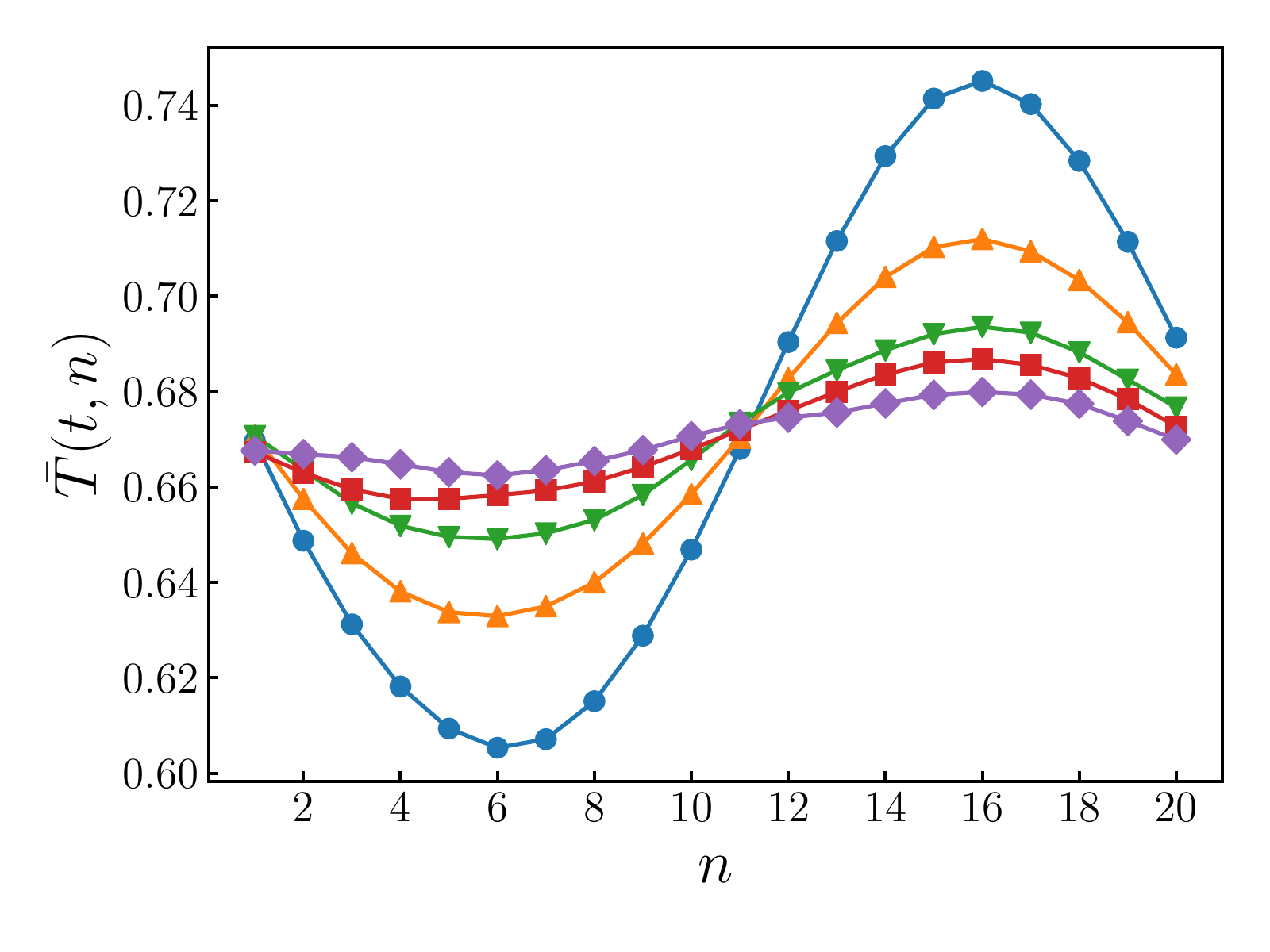}
	\includegraphics[width=0.325\linewidth]{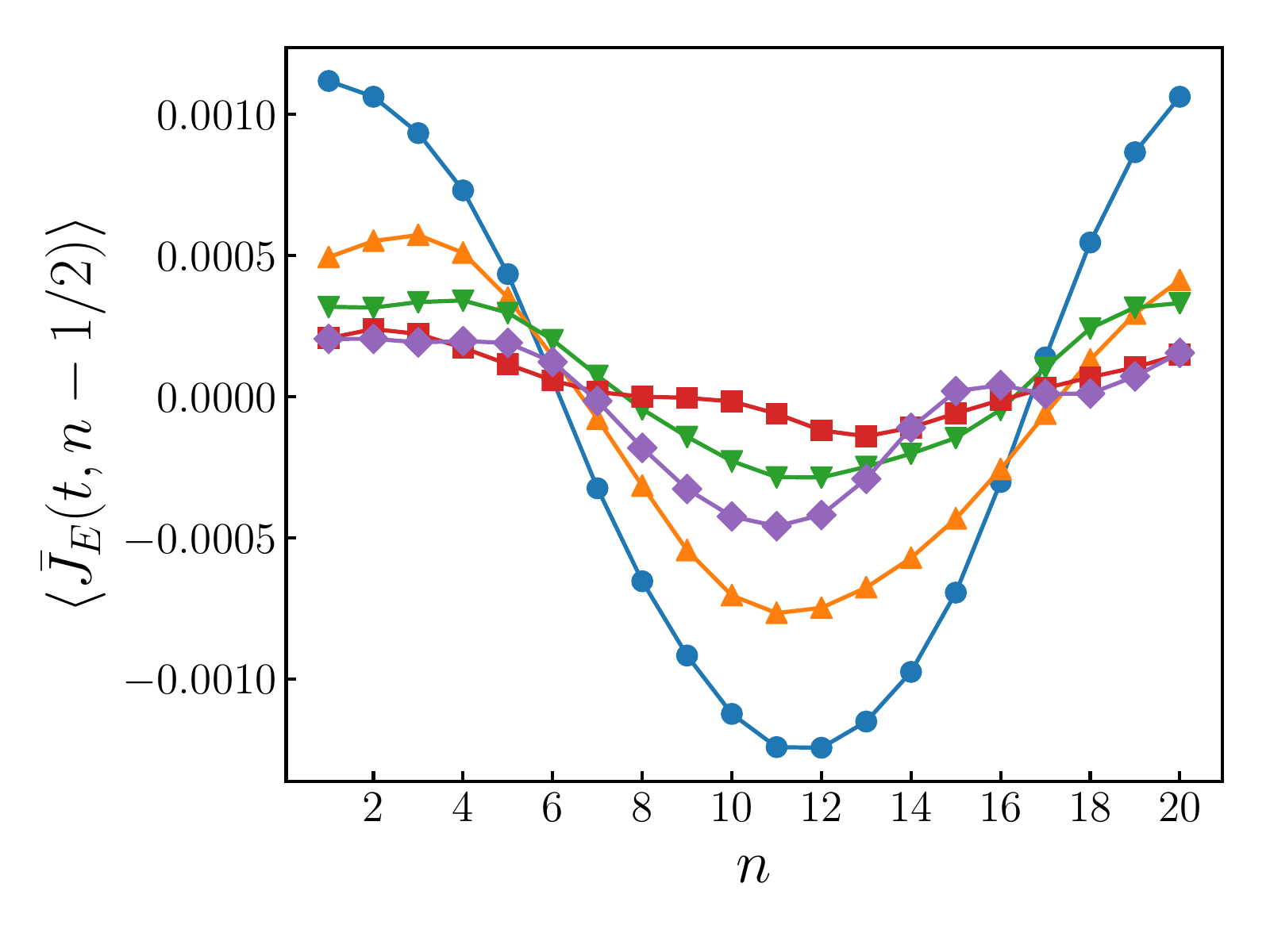}
	\caption{The time evolution of the local energy, the local temperature and the energy current.}
	\label{fig:time_evol}
\end{figure}

\begin{figure}[tbh]
	\centering
	\includegraphics[width=0.325\linewidth]{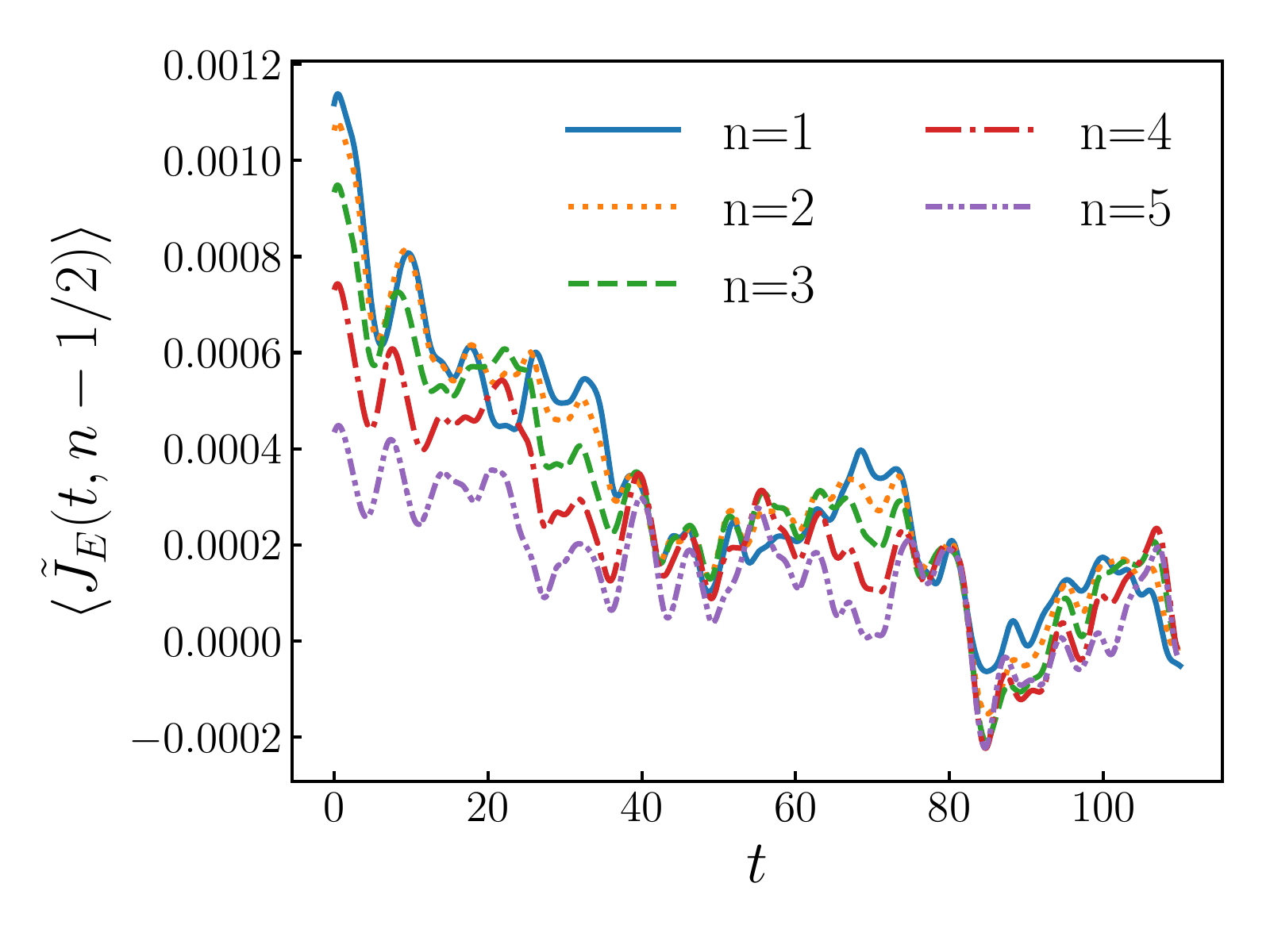} 
	\includegraphics[width=0.325\linewidth]{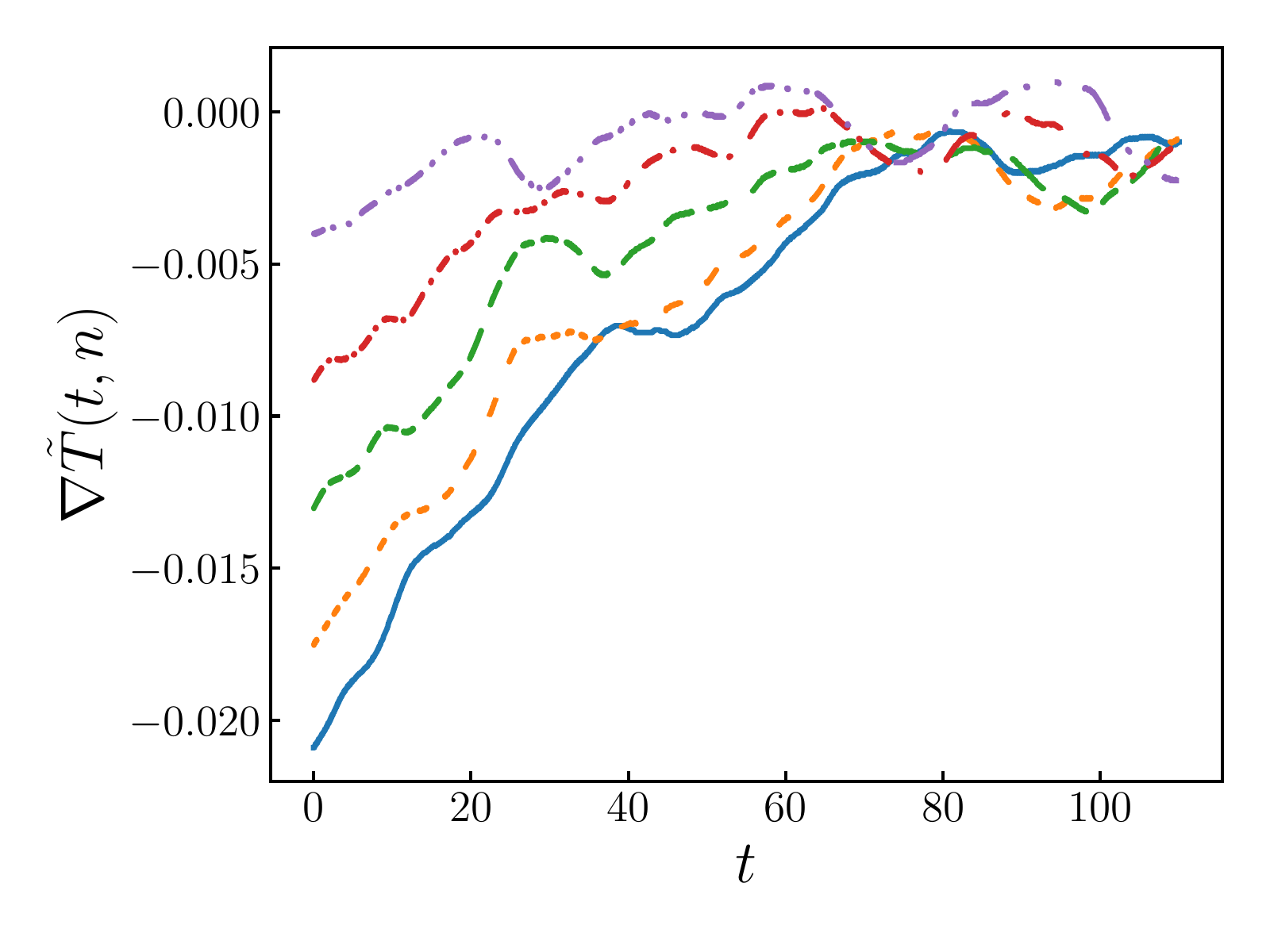}
	\includegraphics[width=0.325\linewidth]{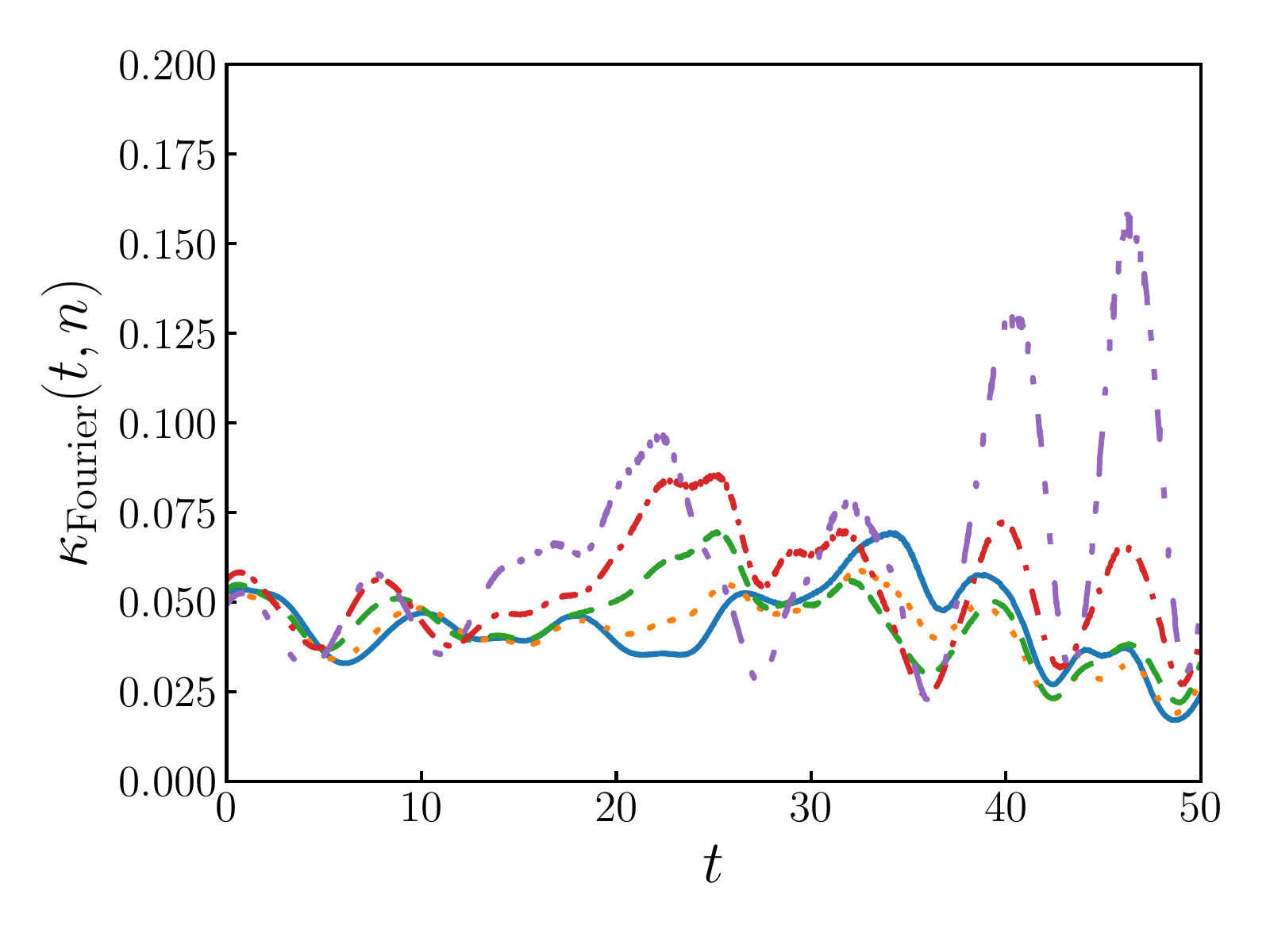}    
	\caption{The time evolution of the energy current, the temperature gradient, and the thermal conductivity at $n=1,\cdots,5$.}
	\label{fig:kfourier}
\end{figure}

If the system is well described by hydrodynamics,
we can extract the thermal conductivity from the Fourier law Eq.~\eqref{eq:CR-heat}.
We introduce temporally averaged operator by 
\begin{align}
    \tilde{O}(t) &\equiv \frac{1}{\tau}\int_{t}^{t+\tau} \diff t' \bar{O}(t').
\end{align}
The time evolution of temporally averaged quantities are shown in Fig.~\ref{fig:time_evol} (and Fig.~\ref{fig:time-ev} in the main text), where $\tau=10$. 
Again, the local temperature is estimated by using the equation of state in equilibrium as $\tilde{T}(t,n) = T(\braket{\tilde{h}(t,n)})$. (In the main text, the tilde is omitted for the simplicity of notation.)
In Fig.~\ref{fig:kfourier}, we show the time evolution of $\nabla \tilde{T}(t,n)$, $\braket{\tilde{J}_E(t,n-1/2)}$ and $\kappa_\text{Fourier}(t,n)$.
We see that the typical equilibration time is $t_\text{eq} \sim 100$
at which there are no net energy current nor temperature gradient.

We remark that evaluation of the thermal conductivity $\kappa_\text{Fourier}(t,n)$ 
from the Fourier law is difficult at a large-time limit $t \geq t_\text{eq}$ or inappropriate site around $n=6$ and $16$.
This is because both temperature gradient $|\nabla \tilde{T}(t,n)|$ and energy current $\braket{\tilde{J}_E(t,n-1/2)}$ takes a small value in those setups 
so that numerical evaluations suffer from large uncertainty.
Thus, we can rigorously estimate the thermal conductivity from the data at $t \ll t_\text{eq}$ and at the node position of the temperature profile.
In the main article, we show the thermal conductivity at $n=1,11$ in Fig.~\ref{fig:heat-conductivity}. 

\begin{figure}[tbh]
	\centering
	\includegraphics[width=0.4\linewidth]{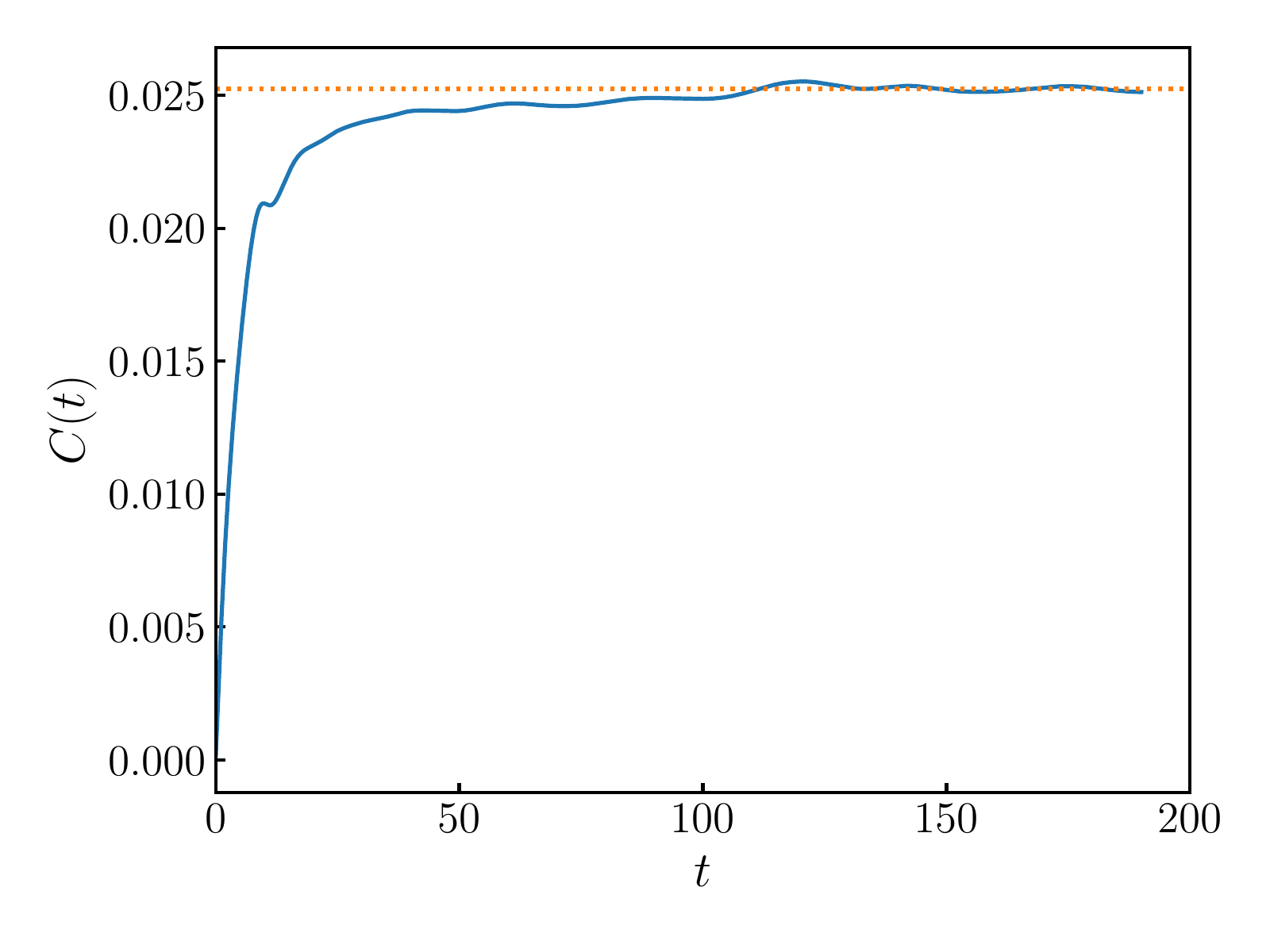}
	\includegraphics[width=0.4\linewidth]{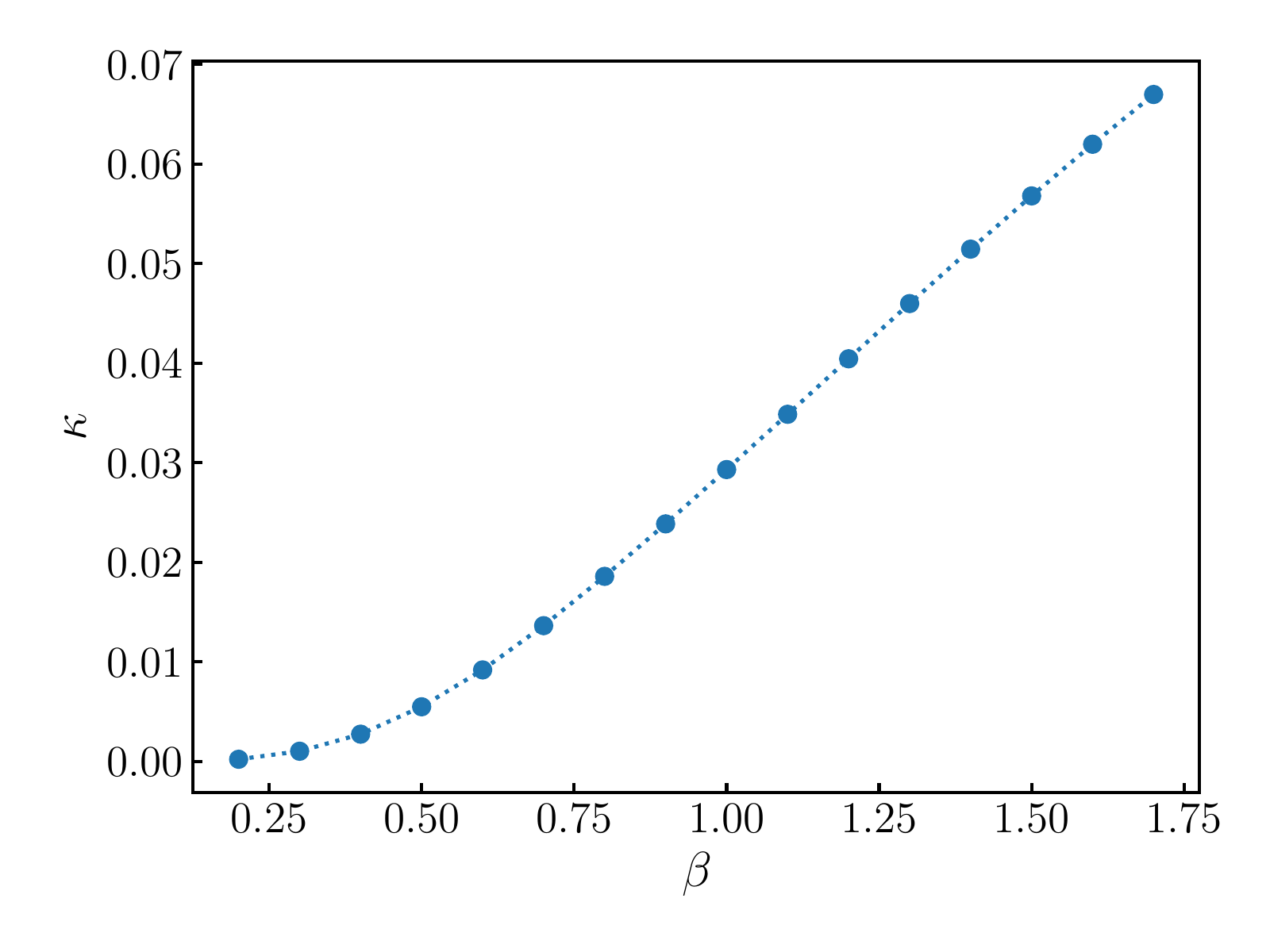}
	\caption{
		(Left) Solid line represents $C(t)$ at $\beta=1.5$. The dotted line shows the late time average of $C(t)$.
		(Right) The $\beta$-dependence of the thermal conductivity.}
	\label{kappaGK}
\end{figure}

\paragraph{Thermal conductivity from the Green-Kubo formula.}
On the basis of the Green-Kubo formula, the thermal conductivity is given by
\begin{align}
	\kappa(t,n)
	= 
	\frac{\beta_n (t)}{4N} \int_{-\infty}^\infty \diff t
	\average{\{\hJsf_E(t), \hJsf_E(0)\} }^{\tpq}_{\beta_n}
	\simeq
	\frac{\beta}{4N} \int_{-\infty}^\infty \diff t
	\average{\{\hJsf_E(t), \hJsf_E(0)\} }^{\tpq}_{\beta}
\end{align}
where $\hJsf_E(t) = \sum_{n=1}^N\hJ^n_E(t)$.
In the second equality, we neglect the spatial and temporal dependence of temperature, which gives rise to the higher-order derivative correction.
Conceptually, the Green-Kubo formula does not require a fully microscopic time resolution but only a hydrodynamic one. 
Therefore, we also replace the bare correlation function $\average{\{\hJsf_E(t), \hJsf_E(0)\} }^{\tpq}_{\beta}$ with the temporally smeared one, which leads to
\begin{align}
G(t) \equiv 
\frac{1}{t_h^2}
\int_t^{t+t_h} \diff s
\int_0^{t_h} \diff s'
\average{\{\hJsf_E(s), \hJsf_E(s')\} }^{\tpq}_{\beta}.
\end{align}
An appropriate time scale $t_h$ is estimated from the damping time of the bare correlation function $\average{\{\hJsf_E(t), \hJsf_E(0)\} }^{\tpq}_{\beta}$. Below, we set $t_h = 10$ which is numerically computed.
The thermal conductivity is now given by
\begin{align}
\kappa(t,n)
\simeq \frac{\beta}{2} \lp C(\infty) - C(-\infty) \rp
 \with
 C(t) 
 = 
 \frac{1}{2N} \int_{0}^t \diff t' G(t').
\end{align}
Since $C(t)$ is defined by the symmetric Green's function, we can show it is equal to $-C(-t)$ within a small error, which is actually confirmed numerically.
Thus, we reach the following simple linear response result 
in the $\ltpq$ formulation:
\begin{align}
\kappa(t,n) \simeq \beta C(\infty) \equiv \kappa
\end{align}
The left panel of Fig.~\ref{kappaGK} demonstrates a typical behavior of the function $C(t)$, from which one can confirm that $C(t)$ converges when $t \gtrsim 100$.
The value at infinitely late time $C(\infty)$ is estimated by the average of $C(t)$ from $t=100$ to $200$.
The resulting thermal conductivity at each $\beta$ is shown in the left panel of Fig.~\ref{kappaGK}, which corresponds to the result shown in Fig.~\ref{fig:heat-conductivity} in the main text.

\subsection{Numerical evaluation of generating functions in fluctuation theorem}

Let us finally present a numerical method to compute the generating functional.
When the Hamiltonian does not depend on time explicitly, 
we can rewrite the generating functions for the total entropy production 
Eqs.~\eqref{eq:GF}-\eqref{eq:GB} as follows:
\begin{align}
G_F^{\ltpq}(z;t) &=
\frac{\rme^{\rmi z \log (Z_{\ltpq}[\{\tilde{\beta}(t,n)\}]/Z_{\ltpq}[\{\tilde{\beta}(0,n)\}])}}{Z_\text{lTPQ}[\{\tilde{\beta}(0,n)\}]}
\Braket{\{\tilde{\beta}(0,n)\}|
	\rme^{\rmi Ht}
	\rme^{\rmi z\hat{K}[\{\tilde{\beta}(t,n)\}]} \rme^{-\rmi \hat{H}t}
	\rme^{-\rmi z\hat{K}[\{\tilde{\beta}(0,n)\}]}
	|\{\tilde{\beta}(0,n)\}}, \\
G_B^{\ltpq}(\rmi -z;t) &=
\frac{\rme^{-\rmi z^*\log(Z_{\ltpq}[\{\tilde{\beta}(t,n)\}]/Z_{\ltpq}[\{\tilde{\beta}(0,n)\}])}}{Z_{\ltpq}[\{\tilde{\beta}(0,n)\}]} \notag \\
&\times
\Braket{\{\tilde{\beta}(t,n)\}|
	\rme^{-\rmi Ht}
	\rme^{-\hat{K}[\{\tilde{\beta}(0,n)\}]}
	\rme^{\rmi\hat{K}[\{\tilde{\beta}(0,n)\}]z^*}
	\rme^{\rmi\hat{H}t}
	\rme^{-\rmi\hat{K}[\{\tilde{\beta}(t,n)\}]z^*}
	\rme^{\hat{K}[\{\tilde{\beta}(t,n)\}]}
	|\{\tilde{\beta}(t,n)\}}^*.
\end{align}
Here $\ket{\{\tilde{\beta}(t,n)\}}$ is an $\ltpq$ state, whose temperature profile is chosen to be that at time $t$.
We compute these quantities based on the following steps:
\begin{enumerate}
	\item Make an $\ltpq$ state $\ket{\{\beta(0,n)\}}$ whose temperature profile is set by the initial condition~\eqref{eq: initial profile}.
	\item Compute $\rme^{-\rmi Ht} \ket{\{\beta(0,n)\}}$.
	\item Compute spatially smeared and temporally averaged local energies $\{\braket{\tilde{h}(t,n)} \}$.
	\item Compute local temperatures $\{\tilde{\beta}(t,n) \}$ using the energy-temperature relation.	\item Make an auxiliary $\ltpq$ state $\ket{\{\tilde{\beta}(t,n)\}}$.
\end{enumerate}
Exponential operators like $\rme^{\rmi z\hat{K}[\{\tilde{\beta}(t,n)\}]}$ and $\rme^{-\rmi z\hat{K}[\{\tilde{\beta}(0,n)\}]}$ are regarded as a time evolution operator.
Action of these operators to pure states are computed by fourth order Runge-Kutta method.

\begin{figure}[tbh]
	\centering
	\includegraphics[width=0.5\linewidth]{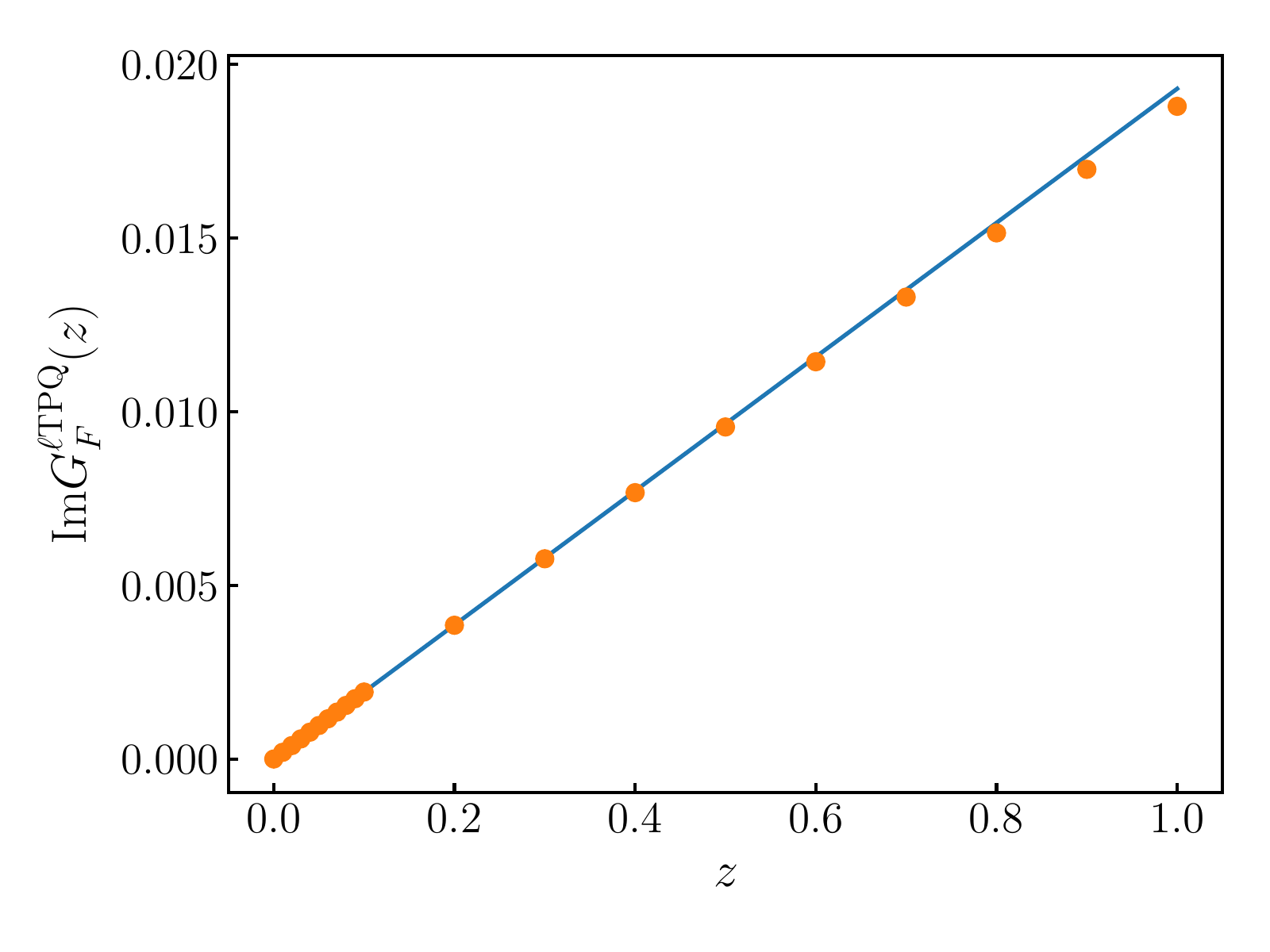}
	\caption{The generating function $G_F^{\ltpq}$ at $t=20$.
	The solid line stands for the fitting function.}
	\label{fig:fit GF}
\end{figure}

The entropy production can be computed by three independent ways.
One way is to compute the expectation value of the entropy operator Eq.~\eqref{entropy} directly. 
The others are to compute the gradient of $\Im G^{\ltpq}_F(z)$ and $\Im G^{\ltpq}_B(\rmi-z)$ at $z=0$.
Concretely, they are given by
\begin{align}
    \Sigma_F \equiv \Im 
 \left. \frac{\diff G_F^{\ltpq}(z)}{\diff (\rmi z)} \right|_{z=0},
    \quad 
    \Sigma_B \equiv \Im 
 \left. \frac{\diff G_B^{\ltpq}(\rmi-z)}{\diff (\rmi z)} \right|_{z=0}.
 \label{eq:Sigma-FB}
\end{align}
In Fig.~\ref{fig:fit GF}, we demonstrate the estimation of the entropy production from $G_F^{\ltpq}$.
At each time, we confirm that the generating function behaves as a linear function around the origin, which allows us to evaluate the entropy production
according to Eq.~\eqref{eq:Sigma-FB}. 
In the figure, the solid line shows the fitting result using the data at $z=0.00, 0.01, \dots, 0.03$. The choice of the fitting range is insensitive to the result. The fitting of $G_B^{\ltpq}(\rmi-z)$ can be performed in the same manner.

\begin{figure}[tbh]
	\centering
	\includegraphics[width=0.5\linewidth]{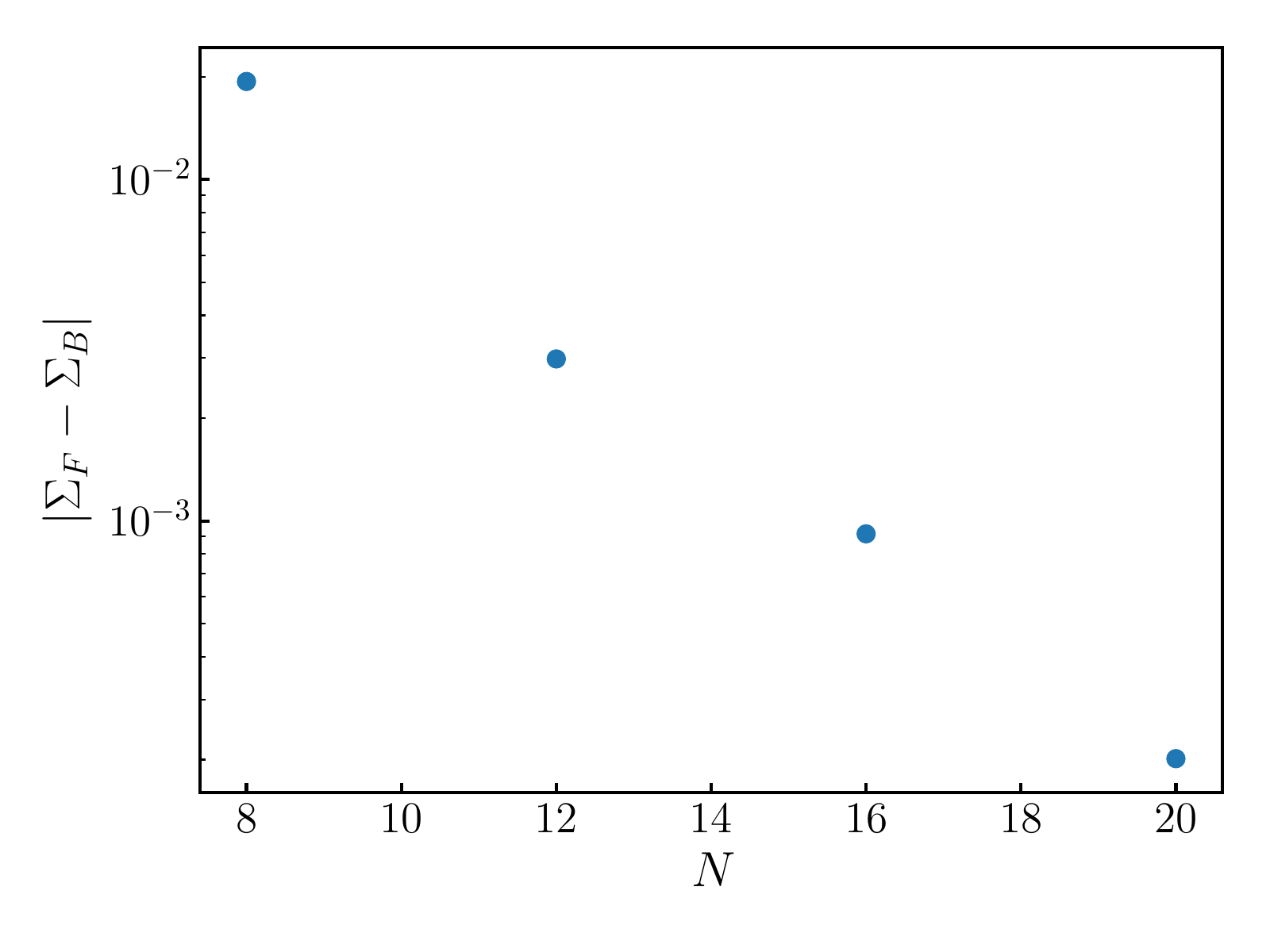}
	\caption{The difference between $\Sigma_F$ and $\Sigma_B$ at $t=20$.}
	\label{fig:dif Sigma FB}
\end{figure}

\begin{figure}[tbh]
	\centering
	\includegraphics[width=0.5\linewidth]{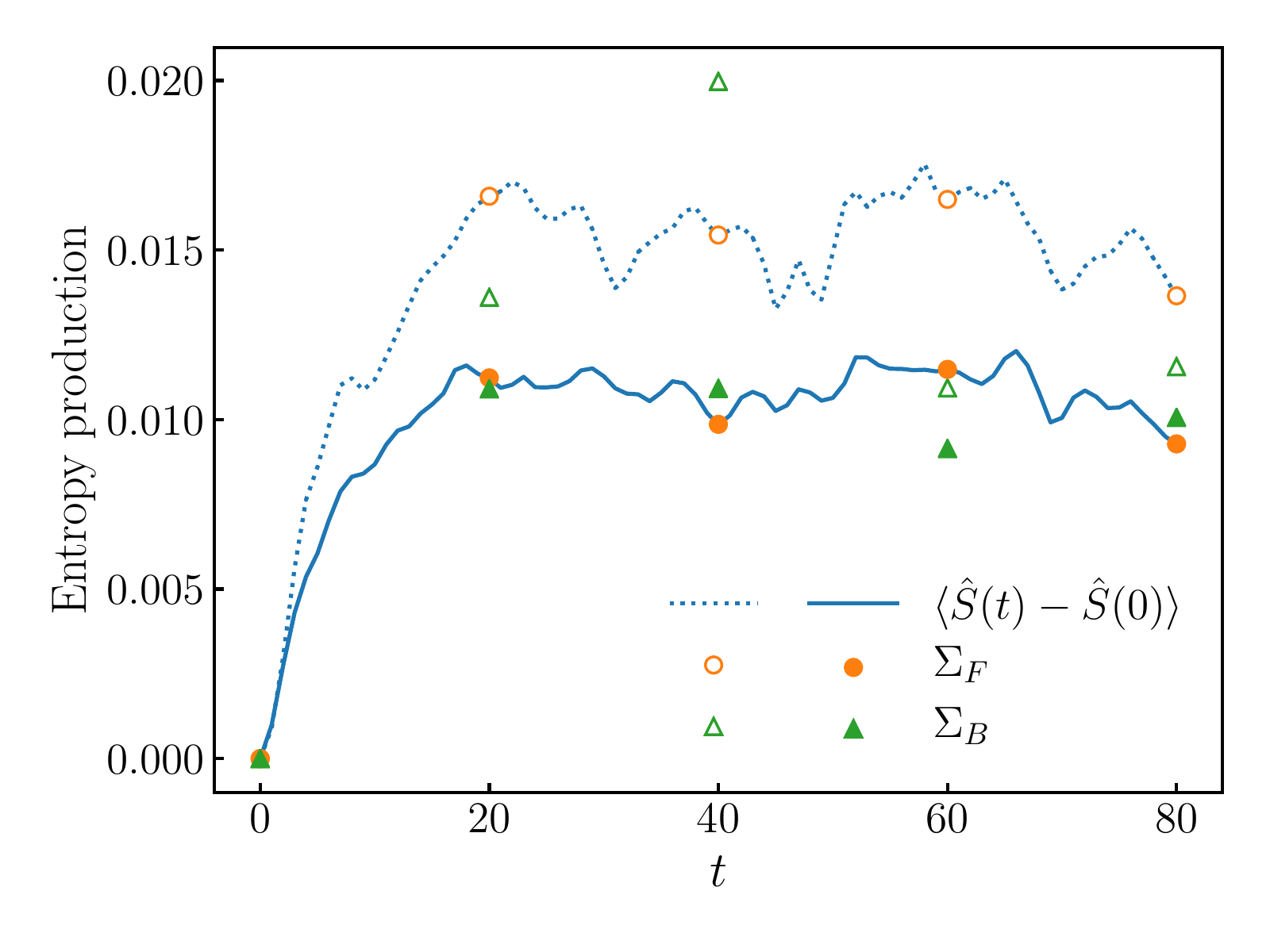}
	\caption{The entropy production for $N=12$. 
	The dotted line and the open symbols show the entropy production obtained for a single random vector. The solid line and the closed symbols show the average over 5 independent samples.}
	\label{fig:entropy seed deps}
\end{figure}

Unlike the standard quantum fluctuation theorem,
$\Sigma_F$ agrees with $\Sigma_B$ only in the large fluid-cell limit.
We numerically confirm that the difference between $\Sigma_F$ and $\Sigma_B$ scales as $\rme^{-N}$, where $N$ is the system size.
In Fig,~\ref{fig:dif Sigma FB}, we show the scaling behavior at $t=20$.
We also check that the difference tends to vanish by taking the random vector average, which is consistent with our proof of equivalence.
We compute the entropy production for 5 random vectors and take the random vector average.
In Fig.~\ref{fig:entropy seed deps}, the solid line and the closed symbols denote the averaged ones while the dotted line and open symbols do the $\ltpq$ result for the entropy production; namely, computed by a single random vector.
Here, numerical simulations are performed for $N=12$.
This result is consistent with the fact that $\Sigma_F$ coincides with $\Sigma_B$ for the standard mixed state (or canonical ensemble) formulation.

\end{document}